\documentclass[fleqn,usenatbib]{mnras}

\usepackage{newtxtext,newtxmath}
\usepackage[T1]{fontenc}

\DeclareRobustCommand{\VAN}[3]{#2}
\let\VANthebibliography\thebibliography
\def\thebibliography{\DeclareRobustCommand{\VAN}[3]{##3}\VANthebibliography}

\usepackage{graphicx}	% Including figure files
\usepackage{amsmath}	% Advanced maths commands

\makeatletter

\patchcmd{\NAT@citex}
  {\@citea\NAT@hyper@{%
     \NAT@nmfmt{\NAT@nm}%
     \hyper@natlinkbreak{\NAT@aysep\NAT@spacechar}{\@citeb\@extra@b@citeb}%
     \NAT@date}}
  {\@citea\NAT@nmfmt{\NAT@nm}%
   \NAT@aysep\NAT@spacechar\NAT@hyper@{\NAT@date}}{}{}

\patchcmd{\NAT@citex}
  {\@citea\NAT@hyper@{%
     \NAT@nmfmt{\NAT@nm}%
     \hyper@natlinkbreak{\NAT@spacechar\NAT@@open\if*#1*\else#1\NAT@spacechar\fi}%
       {\@citeb\@extra@b@citeb}%
     \NAT@date}}
  {\@citea\NAT@nmfmt{\NAT@nm}%
   \NAT@spacechar\NAT@@open\if*#1*\else#1\NAT@spacechar\fi\NAT@hyper@{\NAT@date}}
  {}{}

\makeatother

\title[Learning the properties of the G ring]{Saturn’s G ring: insights from the dynamical evolution of dust particles
from the G-ring arc}

\author[Zhenghan Chen et al.]{
Zhenghan Chen,$^{1,2}$
Xiaodong Liu,$^{1,2}$
and Kun Yang$^{1,2}$\thanks{E-mail: yangk77@mail.sysu.edu.cn}
\\
$^{1}$School of Aeronautics and Astronautics, Shenzhen Campus of Sun Yat-sen University, Shenzhen, Guangdong 518107, China\\
$^{2}$Shenzhen Key Laboratory of Intelligent Microsatellite Constellation, Shenzhen Campus of Sun Yat-sen University, Shenzhen, Guangdong 518107, China\\
}

\date{Accepted XXX. Received YYY; in original form ZZZ}

\pubyear{2015}
\usepackage{lineno}
\let\oldequation\equation
\let\oldendequation\endequation
\renewenvironment{equation}{\linenomathNonumbers\oldequation}{\oldendequation\endlinenomath}
\begin{document}
%\linenumbers
\label{firstpage}
\pagerange{\pageref{firstpage}--\pageref{lastpage}}
\maketitle

\begin{abstract}
%Saturn's G ring is thought to originate from the arc of debris and the moonlet Aegaeon within the arc near the inner edge of the ring.
To explore the formation and properties of Saturn's G ring, we study the dynamics of micron-sized dust particles originating from the arc of debris near the inner edge of the ring.
% origin: To analyze the features of Saturn’s G ring
The dynamical evolution of particles due to various perturbation forces and the plasma sputtering that erodes the particles is simulated by a well-tested numerical code.
Based on the simulation results, the normal $I/F$ of the G ring observed by the Cassini spacecraft can be explained by dust particles originating from the arc.
%is estimated and compared to the observations obtained by the Cassini spacecraft.
%The good agreement in the normal $I/F$ supports that the G ring may originate from the arc (including Aegaeon).
%, where the particles originating from the arc evolve outward due to the plasma drag.
Other properties of the G ring are also estimated, including the steady-state size distribution and the number density of ring particles, 
%the particle properties, 
the geometric optical depth, the apparent edge-on thickness, the age and the remaining lifetime of the G ring.
We find that the particle size distribution of the G ring follows a power law with an exponent of 2.8, and dust particles in the size range of $[5, 10]\,{\mu}$m are dominant within the ring.
% or:Based on the simulation results and the observations obtained by the Cassini spacecraft, the normal $I/F$, the particle properties, the geometric optical depth, the apparent edge-on thickness, the age and the remaining lifetime of the G ring are estimated.
The average number density of particles of the G ring in the radial direction is about $10^{-3}$-$10^{-2}\,\mathrm{m}^{-3}$.
%, and the particle size distribution of the G ring follows a power law with the exponent $q=\textcolor{red}{2.8}$.
%The value of the exponent $q=2.73$ of the particle size distribution of the G ring is constrained.
%By matching our results of the normal $I/F$ of the G ring and the number density near the G ring to the observations obtained by the Cassini spacecraft, the value of the exponent $q=2.73$ of the particle size distribution of the G ring is constrained.}
%Our results of the normal $I/F$ of the G ring and the number density \textcolor{cyan}{near} the G ring match well with the observations obtained by the Cassini spacecraft. 
%The observed normal $I/F$ in the part of the ring that passes through the arc is dominantly contributed by $[5, 20]\,{\mu}$m particles, while that in other parts of the ring are mainly contributed by $[2, 10]\,{\mu}$m particles.
%The particle size distribution of the G ring follows a power law with the exponent $q=2.73$, 
%The peak value of the average particle number density of the G ring is on the order of $10^{-2}\,\mathrm{m}^3$.
%In addition to the G ring, dust particles originating from the arc can also be detected in the region between Saturn's main ring and the G ring. 
The peak value of the edge-on geometric optical depth of the G ring is about $3.9\times10^{-2}$. The maximum apparent edge-on thickness of the G ring with the geometric optical depth larger than $1\times10^{-8}$ is approximately $9,000\,\mathrm{km}$. The age of the G ring is estimated to be $10^{6}$-$10^{7}\,\mathrm{years}$, and the remaining lifetime of the ring is on the order of $10^{4}\,\mathrm{years}$.
\end{abstract}


%%%%%%%%%%%%%%%%%%%%%%%%%%%%%%%%%%%%%%%%%%%%%%%%%%

%%%%%%%%%%%%%%%%% BODY OF PAPER %%%%%%%%%%%%%%%%%%

\section{Introduction}
Saturn's G ring is a faint ring composed of micron-sized dust particles, which is located approximately $166,000$ to $175,000\,\mathrm{km}$ from Saturn.
The ring was first imaged by the Voyager 1 in 1980 \citep{smith1981encounter}. 
Subsequently, the investigation of the G ring can be divided into two periods: before and after the arrival of the Cassini spacecraft.
In the first epoch, several studies \citep{van1983absorption, showalter1993seeing, canup1997evolution, meyer1998constraints, throop1998g, lissauer2000hst} were focused on the structure and particle properties of the G ring. However, most of these studies were carried out by analyzing the observation data, such as the analyses of the proton absorption signature detected by the Pioneer 11 \citep{van1983absorption}, images obtained by the Voyager 1 and 2 \citep{showalter1993seeing}, dust impact signals \citep{meyer1998constraints} and images obtained by the Hubble Space Telescope \citep{lissauer2000hst}.
Less attention was paid to the dynamics of dust particles within the ring.
%, which determines the structure and the particle properties of the ring.Only the model that tracks the number density evolution of the G ring used by \citet{canup1997evolution} and \citet{throop1998g} had considered the plasma drag.
%\textcolor{red}{In addition, some properties of the G ring were still unclear, such as the edge-on geometric optical depth and the thickness of the G ring.}
%In addition, some properties of the G ring were not strongly constrained in previous studies, such as the edge-on geometric optical depth and the thickness of the G ring.

After the arrival of the Cassini spacecraft, a bright arc (hereafter refer as "G-ring arc") and a moonlet Aegaeon that orbits within the arc were discovered near the inner edge of the G ring \citep{hedman2007source,hedman2010aegaeon}.
Scientists began to take an interest in the dynamics of dust particles in order to explore the relationships among the three components of the G-ring system, namely Aegaeon, the arc, and the ring.
\citet{madeira2018production} and \citet{madeira2020effects} investigated the roles of Aegaeon in the production and fate of particles within the G-ring arc by analyzing the dynamical evolution of particles within the arc.
They showed that many particles confined in the arc collide with Aegaeon, and Aegaeon is not the only source of particles within the arc.
% However, the dynamics of particles after they leave the arc were paid little attentions to by \citep{madeira2018production} and \citep{madeira2020effects}, which may form the G ring as suggested by \citep{hedman2007source}.
To investigate the formation of the G ring, \citet{maravilla2014saturnian} used an analytical model to study the dynamics of particles ejected from Aegaeon, and suggested that Aegaeon can supply sub-micron and nanometric dust particles to the G ring.
% that considered the Lorentz force and gravity of Saturn
Except for Aegaeon, \citet{hedman2007source} proposed that the G-ring arc contains a population of large debris in the size range from centimeters to meters, which also supply dust particles to the G ring.
%However, \citet{maravilla2014saturnian} paid little attention to particles originating from the arc, which may \textcolor{red}{escape from the arc and} contribute to the G ring.
The study on the dynamics of particles within analogous dynamical systems, i.e.~the arcs of Saturn's moons Anthe and Methone, showed that particles can escape from the arc due to the plasma drag \citep{sun2017dust}. This mechanism may also result in the ejection of particles from the G-ring arc to the G ring.

In this paper, the dynamics of micron-sized dust particles originating from the G-ring arc is studied by high-accuracy numerical simulations to explore the formation and properties of the G ring.
%learn the features of the G ring.
Various perturbation forces and the plasma sputtering are taken into account.
The dynamical model is described in Sections.~\ref{Dynamical model}, and the numerical simulation strategy is given in Sections.~\ref{Numerical simulation}.
In Section.~\ref{Simulation result}, the orbital evolution, the average dynamical lifetimes and the final fates of particles with different sizes are investigated.
Based on the study of the dynamics of dust particles and the observations obtained by the Cassini spacecraft, the normal $I/F$, the particle properties, the edge-on geometric optical depth, the apparent edge-on thickness, the age and the remaining lifetime of the G ring are estimated and analyzed in Section.~\ref{Properties of the G ring}.
Finally, the conclusion of this paper is given in Section.~\ref{Conclusions}.
%From these results, the apparent edge-on thickness of the G ring is analyzed, and the age and the remaining lifetime of the G ring are estimated.

\section{Dynamical model}
\label{Dynamical model}
Micron-sized dust particles can be generated through the mutual collision between the large debris within the G-ring arc or by the impact of micrometeoroids onto these large bodies. After dust particles start their lives, they are subjected to various perturbation forces, including the effect of planetary oblateness, solar radiation pressure, Poynting-Roberson drag, Lorentz force, plasma drag and the gravity of moons. The equation of dust grain's motion in the equatorial inertial frame of Saturn (SIF), where the $x$-axis of SIF points toward the ascending node of Saturn’s orbit at J2000, the $z$-axis is along the spin axis of Saturn, and the $y$-axis is determined by the right-hand rule, is expressed as
\begin{equation}
    \begin{split}
\ddot{\vec{r}}&=GM_\mathrm{s}\triangledown\left\lbrace\frac{1}{r}\left[1-\sum_{n=1}^{N_S}J_{2n}\left(\frac{R}{r}\right)^{2n}P_{2n}(\cos{\theta})\right]\right\rbrace\\
&+\frac{3S_\mathrm{flux}Q_\mathrm{pr}}{4\rho_\mathrm{g}r_\mathrm{g}c}\left\lbrace\left[1-\frac{(\Dot{\Vec{r}}-\Dot{\Vec{r}}_\mathrm{sun})\cdot\Vec{S}}{c}\right]\Vec{S}-\frac{\Dot{\Vec{r}}-\Dot{\Vec{r}}_\mathrm{sun}}{c}\right\rbrace \\
&+\frac{3\epsilon_0\Phi}{\rho_\mathrm{g}r_\mathrm{g}^2}(\Vec{V}_\mathrm{rel}\times\Vec{B})-\frac{3N_{\mathrm{W}^+}m_{\mathrm{W}^+}|\Vec{V}_\mathrm{rel}|}{4\rho_\mathrm{g}r_\mathrm{g}}\Vec{V}_\mathrm{rel}\\
&-\sum\left[\frac{GM_\mathrm{moons}}{{|\Vec{r}-\Vec{r}_\mathrm{moons}|}^3}(\Vec{r}-\Vec{r}_\mathrm{moons})+\frac{GM_\mathrm{moons}}{|\Vec{r}_\mathrm{moons}|^3}\Vec{r}_\mathrm{moons}\right],
    \end{split}
    \label{eq:dynamic}
\end{equation}
where $\Vec{r}$ is the radius vector of the grain, $M_\mathrm{s}$ the mass of Saturn, $R$ the reference radius for gravitational field of Saturn, $J_{2n}$ and $P_{2n}$ the Saturn’s zonal harmonics and the Legendre functions of degree $2n$, and $\theta$ the colatitude of the grain in the Saturn-centered body-fixed frame.
The parameters of Saturn gravity field are taken from \citet{jacobson2022orbits}, where the $J_2$, $J_4$ and $J_6$ are considered here, i.e., $N_S = 3$.
The variable $S_\mathrm{flux}$ is the solar flux at the distance of Saturn's orbit around the Sun, and $Q_\mathrm{pr}$ is the solar radiation pressure efficiency calculated by Mie theory \citep{burns1979radiation}.
The bulk density of the grain $\rho_\mathrm{g} = 540\,\mathrm{kg\,m^{-3}}$ is assumed to be consistent with that of Aegaeon, the value of which is taken from \citet{thomas2013inner}.
The variable $r_\mathrm{g}$ is the grain size, $c$ the speed of light, $\Vec{r}_\mathrm{sun}$ the radius vector of the Sun, and $\Vec{S}$ the unit vector from the Sun to the grain.

% For simplicity, $\Phi$ is assumed to be a constant.
% The impact of this assumption\textcolor{red}{simplicity} is small, because the variation of dust grains' potential only affects the Lorentz force on them, and small particles which are strongly affected by Lorentz force have a short lifetime (see Sect.~\ref{Simulation result}).
The variable $\epsilon_0$ is the vacuum permittivity, and $\Vec{V}_\mathrm{rel}$ is the velocity of the grain relative to the corotation velocity which is calculated by $\Vec{V}_\mathrm{corotate} = \Omega_\mathrm{s}\times\Vec{r}$, where $\Omega_\mathrm{s}$ is the spin rate of Saturn.
As suggested by \citet{wahlund2005inner}, the electric potential $\Phi$ of the surface of dust grain in the G ring is close to that of the Cassini spacecraft when it orbits above the G ring, where the value is about $-2\,\mathrm{V}$. 
The coefficient of Saturn's magnetic field $\vec{B}$ is adopted from \citet{dougherty2018saturn}.
%The plasma in the region of the G ring was found to be mainly composed of water group ions $\mathrm{W}^+$ and molecular oxygen ion $\mathrm{O}_2^+$ \citep{tokar2005cassini}.
%The densities of these ions are shown to vary yearly in the \textcolor{cyan}{recent} studies \citep{elrod2012seasonal,elrod2014seasonal}.
%Since the observation data of the G ring that we use to constrain our results of the properties of the G ring (see Sect.~\ref{Properties of the G ring}) are obtained in 2006 \citep{hedman2007source} and 2008 \citep{ye2016situ}, and the density ratio of $\mathrm{O}_2^+$ to $\mathrm{W}^+$ is very low in these years, we only consider the plasma drag due to the $\mathrm{W}^+$ here.
The plasma distribution in the inner magnetosphere of Saturn derived by \citet{persoon2020distribution} is used to calculate the plasma drag. The variable $N_{\mathrm{W}^+}$ is the number density of water group ions $\mathrm{W}^+$.
%origin:Here we only consider the plasma drag due to the water group ions $\mathrm{W}^+$ since it contributed the most to the plasma in the vicinity of the G ring in the years of obtaining the observation data \citep{elrod2012seasonal,elrod2014seasonal,hedman2007source,ye2016situ}, which are used to constrain our results of the properties of the G ring (see Section.~\ref{Properties of the G ring}). The number density $N_{W^+}$ of water group ions is derived by \citet{persoon2020distribution}.
The average mass $m_{\mathrm{W}^+}$ is $17.5\,\mathrm{amu}$, and the velocity of the plasma is assumed to equal the corotation velocity.
The variables $M_\mathrm{moons}$ and $\Vec{r}_\mathrm{moons}$ are the mass and the radius vector of moons.
The gravitational perturbations induced by Mimas, Enceladus, Tethys and Aegaeon are considered.

In addition to the perturbation forces, the plasma sputtering which leads to the erosion of dust particles is also taken into account in the simulations. 
The relationship between the sputtering yield (and thus the sputtering rate, as shown in \citet{johnson2008sputtering}) and the grain size has been investigated by \citet{jurac2001saturn}, which showed that the sputtering yield is approximately the same for particles with different %grain
sizes when the incident ion energy is less than 1 keV. Because the energy of the plasma in the vicinity of the G ring is smaller than 100 eV \citep{tseng2013modeling}, in our work the sputtering rate is assumed to be the same regardless of the particle's initial size in the simulations.
Additionally, since the locations of Saturn’s E and G rings are close to each other, the sputtering rate is assumed to be the same as that of the E ring particles in our simulation, which reads \citep{jurac2001saturn,juhasz2002saturn}
\begin{equation}
    r_\mathrm{g}(t_y)=r_\mathrm{g_0}-\frac{t_y}{50}\times10^{-6}
    \label{eq:sputtering},
\end{equation}
where $r_\mathrm{g_0}$ is the initial grain radius in meters, and $t_y$ is the time in years from the moment when the particle is generated from the arc.
%\textcolor{blue}{Note that the sputtering rate is approximately the same for particles with different sizes due to the small energy of the plasma in the vicinity of the G ring \citep{jurac2001saturn,johnson2008sputtering,tseng2013modeling}.}

By simulating the dynamics of test particles, it is found that dust particles from the G-ring arc spend only a small percent of their lifetimes in Saturn’s shadow, and their dynamical evolutions are rarely affected.
Therefore, the planetary shadow effects are neglected in simulations of this work.
%Note that the criterion for detecting whether the particle is in the shadow is adopted from \citet{liu2016dynamics}.
%Note that the charging processes which is important for depending the Lorents force is not taken into account in this work, since the small particles which are most affected by the Lorentz force has a short lifetime (see Section 3).

\section{Numerical simulation Strategy}
\label{Numerical simulation}
\begin{figure}
        \includegraphics[width=\columnwidth]{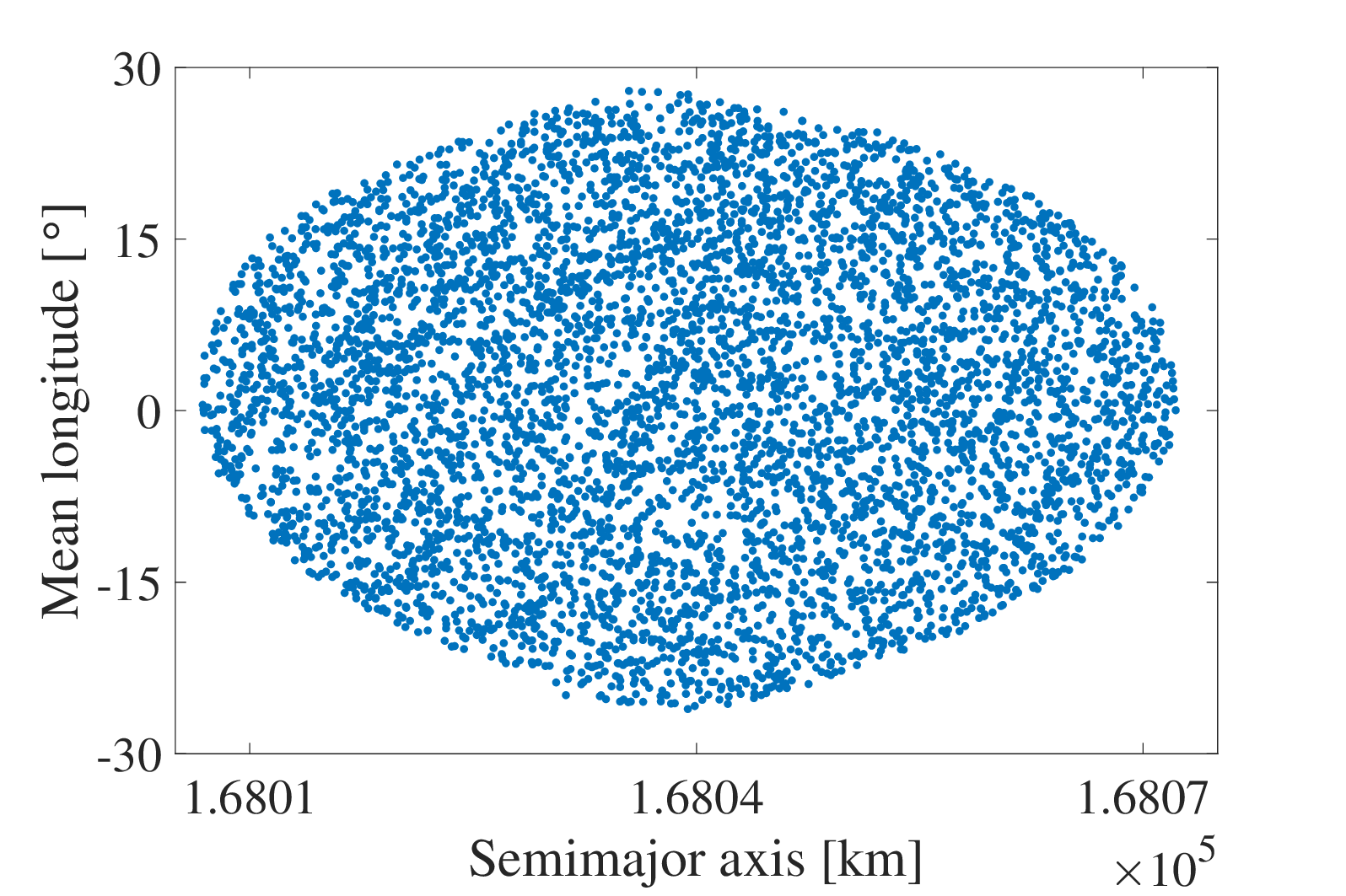}
    \caption{Initial semi-major axes and mean longitudes of the simulated particles.}
    \label{initial_element}
\end{figure}

%Dust particles start their lives within the G-ring arc as they are generated from the large debris in the arc.
The G-ring arc and Aegaeon are trapped in a 7:6 corotation eccentricity resonance (CER) with Mimas \citep{hedman2007source, hedman2010aegaeon}. The arc is longitudinally confined in $60^\circ$ and the maximum amplitude of the semimajor axes libration of debris inside the arc is approximately $31\,\mathrm{km}$ \citep{hedman2007source}.
%which implies that their semimajor axes are confined within the range of $[a_c-Wc/2, a_c+Wc/2]$, where $a_c$ is the central semimajor axis of the resonance and $W_c$ is the resonance width of $64\,\mathrm{km}$, and their longitudes are confined within a interval of $60^\circ$ \citep{madeira2020effects}.
Meanwhile, Aegaeon stays close to the center of the resonance, where the amplitude of its semimajor axis libration is only a few kilometers \citep{hedman2010aegaeon}.
%with its semimajor axis libration amplitude of $4\,\mathrm{km}$.
%, where $a$ is the semimajor axis and $\lambda$ is the mean longitude.
%The intervals of $a$ and $\lambda$ covered by the lobe are the resonance width of $64\,\mathrm{km}$ and $60^\circ$, respectively.
As dust particles are generated from the large debris in the arc, the mean longitudes $\lambda$ of grains are selected to be randomly distributed within $30^\circ$ from the arc's center, and the semimajor axes $a$ are selected to be randomly distributed within $31\,\mathrm{km}$ from Aegaeon’s semimajor axis.
%and the initial mean longitudes $\lambda$ are selected randomly within $30^\circ$ from the arc's center in our simulations.
In this way, 1000 particles are selected for each grain size, from which we exclude the particles initially outside the resonance \citep{madeira2020effects}, and as a result more than 600 grains per size are simulated finally in our simulation.
The initial semimajor axes and mean longitudes of the simulated particles are shown in Fig.~\ref{initial_element}.
Other initial orbital elements of dust particles, including eccentricity $e$, inclination $i$, longitude of ascending node $\Omega$ and argument of pericenter $\omega$, are simply assumed to be the same as those of Aegaeon.

The dynamical evolution of dust grains with 9 different grain sizes, including $0.1\,{\mu}$m, $0.2\,{\mu}$m, $0.5\,{\mu}$m, $1\,{\mu}$m, $2\,{\mu}$m, $5\,{\mu}$m, $10\,{\mu}$m, $20\,{\mu}$m, and $50\,{\mu}$m, is simulated by a well-tested code \citep[see][]{liu2016dynamics,liu2018dust,liu2021configuration,yang2022distribution,chen2024life}. 
The simulation stops if the dust grain hits a moon, reaches the outer edge of the A ring ($\sim$2.27 equatorial radius of Saturn $R_\mathrm{s}$), moves outside the orbital distance of Enceladus ($\sim$$3.95\,R_\mathrm{s}$), or if the grain size decreases to $0.03\,{\mu}$m due to the plasma sputtering.

\section{Dynamical evolution of dust particles}
\label{Simulation result}
From our simulation results, three typical orbital evolutions of dust particles originating from the G-ring arc are found, as shown in Fig.~\ref{orbital evolution}.
For dust particles smaller than $1\,{\mu}$m, both of their semimajor axes and eccentricities increase to large values quickly (within a few years) due to the strong effects of perturbation forces, including the Lorentz force, solar radiation pressure and plasma drag.
However, their orbital distances from Saturn decrease slowly with time (see Fig.~\ref{orbital evolution}a).
%However, \textcolor{red}{their orbital distance shows a trend of gradual decrease.czh
The grain size of $1\,{\mu}$m seems to be a critical size. For particles with this critical size, their semimajor axes generally increase much slower than that of particles smaller than $1\,{\mu}$m, %and the eccentricities of these particles librate and reach a value larger than 0.1 in a few years.
%As a result, the orbital distances of $1\,{\mu}$m dust particles oscillate with large amplitudes within a few years. 
and their orbital distances oscillate with large amplitudes within a few years due to the large libration of their eccentricities (see Fig.~\ref{orbital evolution}b). %czh
\begin{figure}
        \includegraphics[width=\columnwidth]{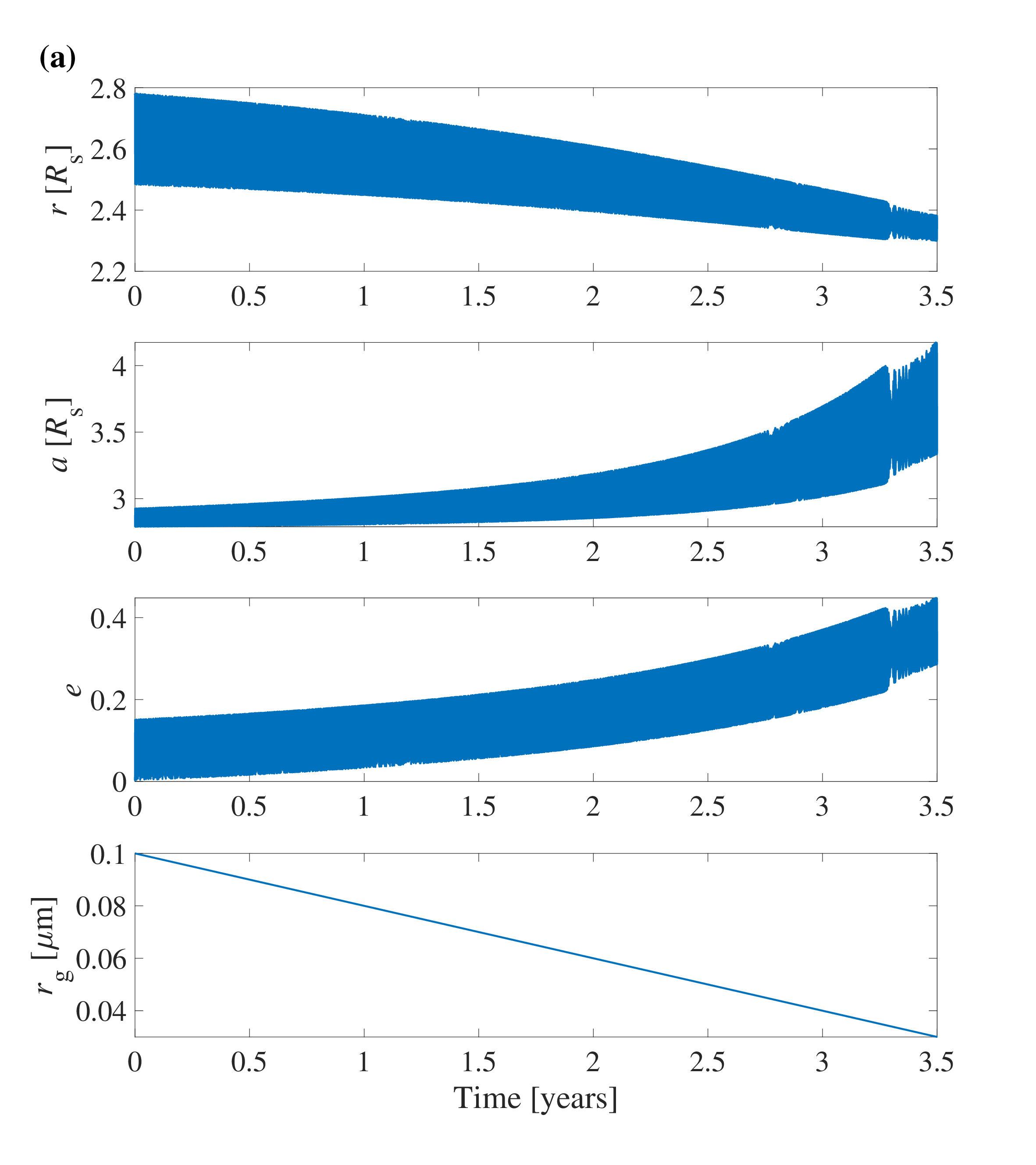}
\end{figure} 
\begin{figure}
        \includegraphics[width=\columnwidth]{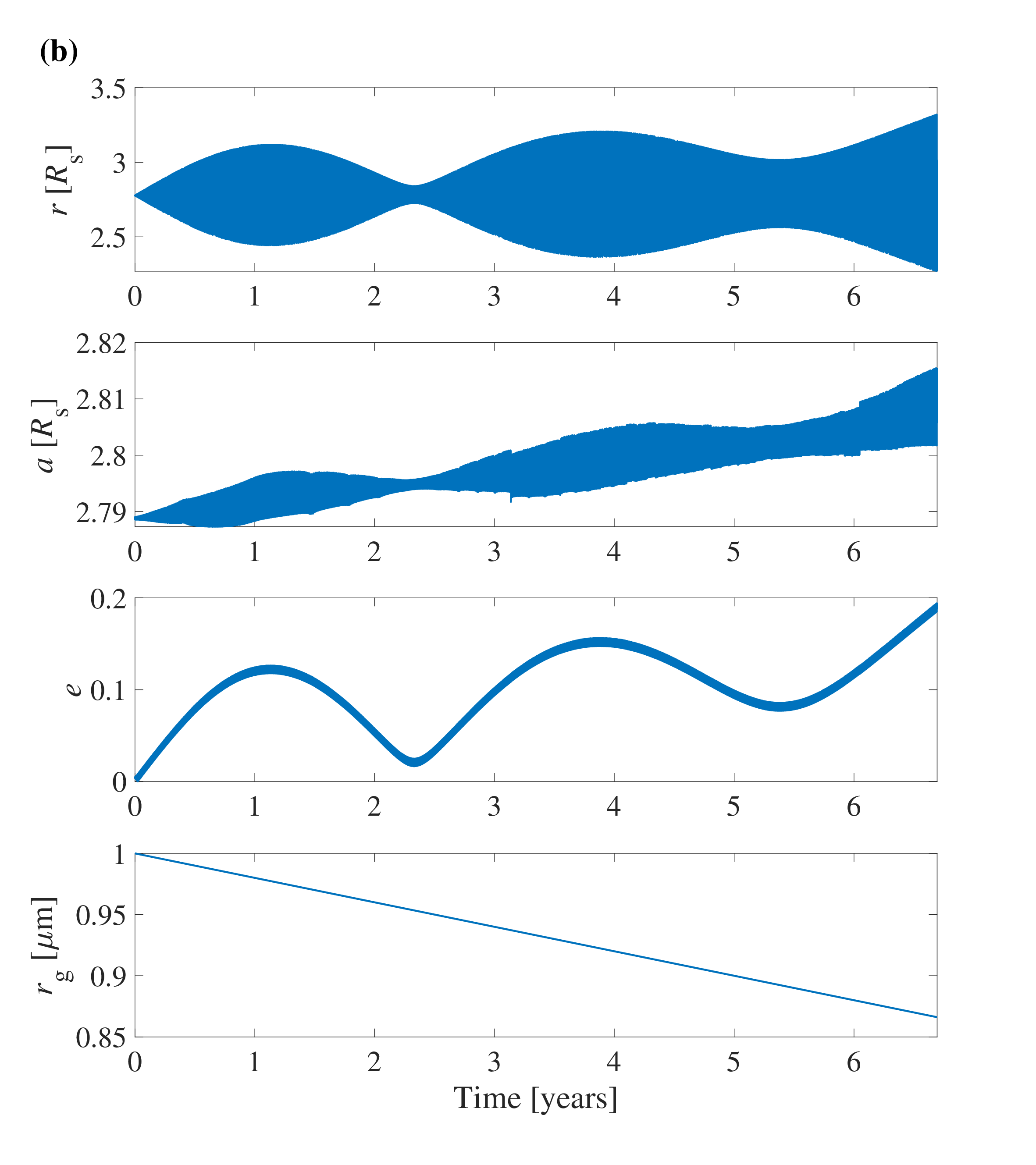}
\end{figure}
\begin{figure}
        \includegraphics[width=\columnwidth]{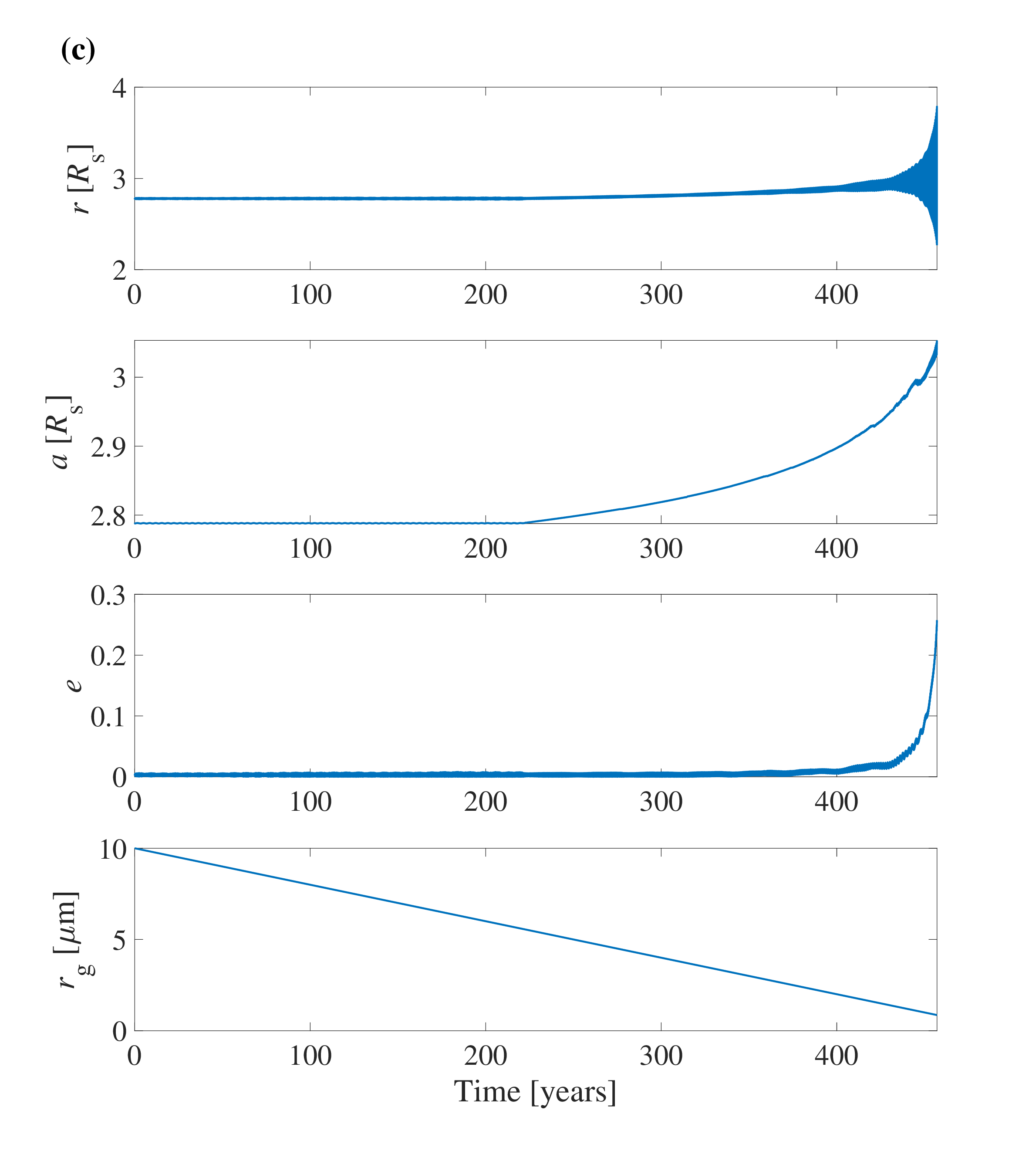}
    \caption{\textit{Panel a:} typical orbital evolution, including the orbital distance, semimajor axis and eccentricity, and physical radius evolution of particle with grain size $r_\mathrm{g}<1\,{\mu}$m. \textit{Panel b:} same as \textit{Panel a}, but with grain size $r_\mathrm{g}=1\,{\mu}$m. \textit{Panel c:} same as \textit{Panel a}, but with grain size $r_\mathrm{g}>1\,{\mu}$m.}
    \label{orbital evolution}
\end{figure}

\begin{figure}
        \includegraphics[width=\columnwidth]{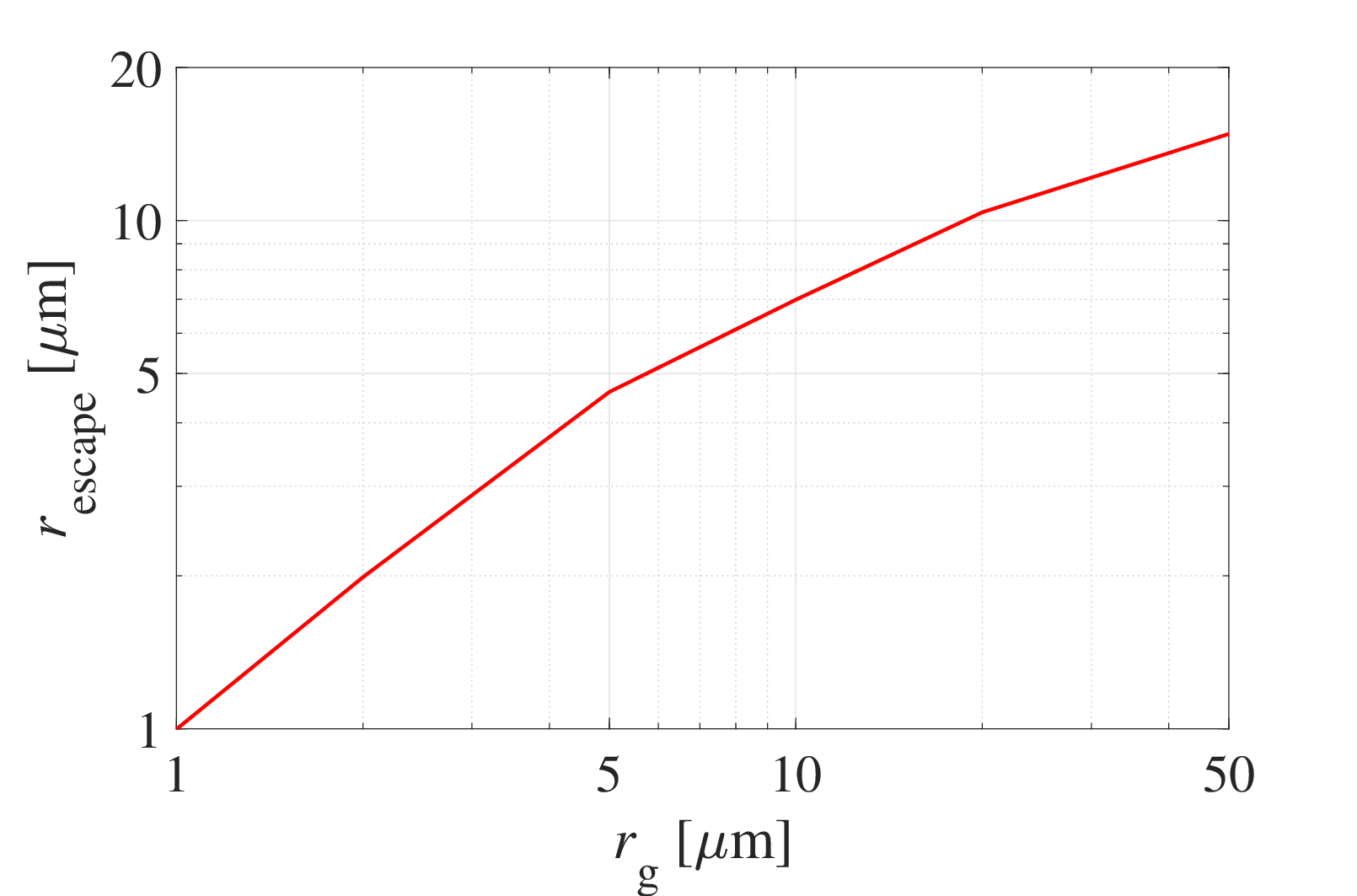}
    \caption{Average grain sizes when dust particles escape from the arc as a function of the initial grain size.}
    \label{r_escape}
\end{figure}

For particles larger than $1\,{\mu}$m, their semimajor axes barely change at first since they are trapped in the arc.
After a period of time, particles can escape from the arc due to the plasma drag, and their semimajor axes increase slowly with time after leaving the arc.
Note that the particle is considered to escape from the arc once its semimajor axis is larger than $62\,\mathrm{km}$ away from the semimajor axis of Aegaeon.
We calculate the average sizes $r_\mathrm{escape}$ of these particles at the moment when they escape from the arc (Fig.~\ref{r_escape}).
%for these particles with \textcolor{blue}{initial} size larger than $1\,{\mu}$m (Fig.~\ref{r_escape}).
Generally, dust particles leave the arc with their grain sizes less than about $16\,{\mu}$m.
%, which is close to the center of the resonance \citep{hedman2010aegaeon}. Here $64\,\mathrm{km}$ is the resonance width of the 7:6 CER \citep{madeira2020effects}.
The eccentricities of these particles remain small during most of their dynamical lifetime, and increase rapidly in the rest of their lifetime due to the strong effects of the Lorentz force and solar radiation pressure.
%\textcolor{red}{This is because the Lorentz force and solar radiation pressure become the dominant perturbation forces only when the grain size of particles become small.}
The orbital distances of these particles remain almost constant during the period when the eccentricities remain small and oscillate significantly in the rest of their lifetime due to the rapid growth of the eccentricities (see Fig.~\ref{orbital evolution}c).

%For dust particles with \textcolor{blue}{initial} sizes in the range of $[1, 5]\,{\mu}$m, there is only a slight decrease in their grain size before they leave the arc; while for dust particles \textcolor{red}{initially} larger than $5\,{\mu}$m, \textcolor{red}{they are trapped in the arc until their grain sizes decrease to the range of $[5, 15]\,{\mu}$m. For particles with initial size of $50\,{\mu}$m, the average grain size when they escape from the arc is about $15\,{\mu}$m.}
%\textcolor{red}{, i.e., when the semimajor axis of the particle exceeds Aegaeon's by more than $64\,\mathrm{km}$,} for particles with \textcolor{blue}{the initial} size larger than $1\,{\mu}$m (Fig.~\ref{r_escape}).
%they are trapped in the arc until their grain sizes decay to nearly $5\,{\mu}$m regardless of their initial grain size.
%\textcolor{red}{Additionally, }the average sizes $r_\mathrm{ecc}$ of these particles ($r_\mathrm{g}>1\,{\mu}$m) at the moment when their eccentricities begin to increase rapidly\textcolor{red}{, which is assumed to be the moment that the eccentricity equals 0.05 (see Fig.~\ref{orbital evolution}c),} are also calculated (Fig.~\ref{r_egetlarge}).
%\begin{figure}
%        \includegraphics[width=\columnwidth]{r_ecc.eps}
%    \caption{Average grain sizes when dust particles' eccentricities begin to increase rapidly as a function of \textcolor{blue}{the initial} grain size.}
%    \label{r_egetlarge}
%\end{figure}
%All the particles' eccentricities begin to increase rapidly when their sizes decay to below $\textcolor{red}{3}\,{\mu}$m or even approach $1\,{\mu}$m.

\begin{figure}
        \includegraphics[width=\columnwidth]{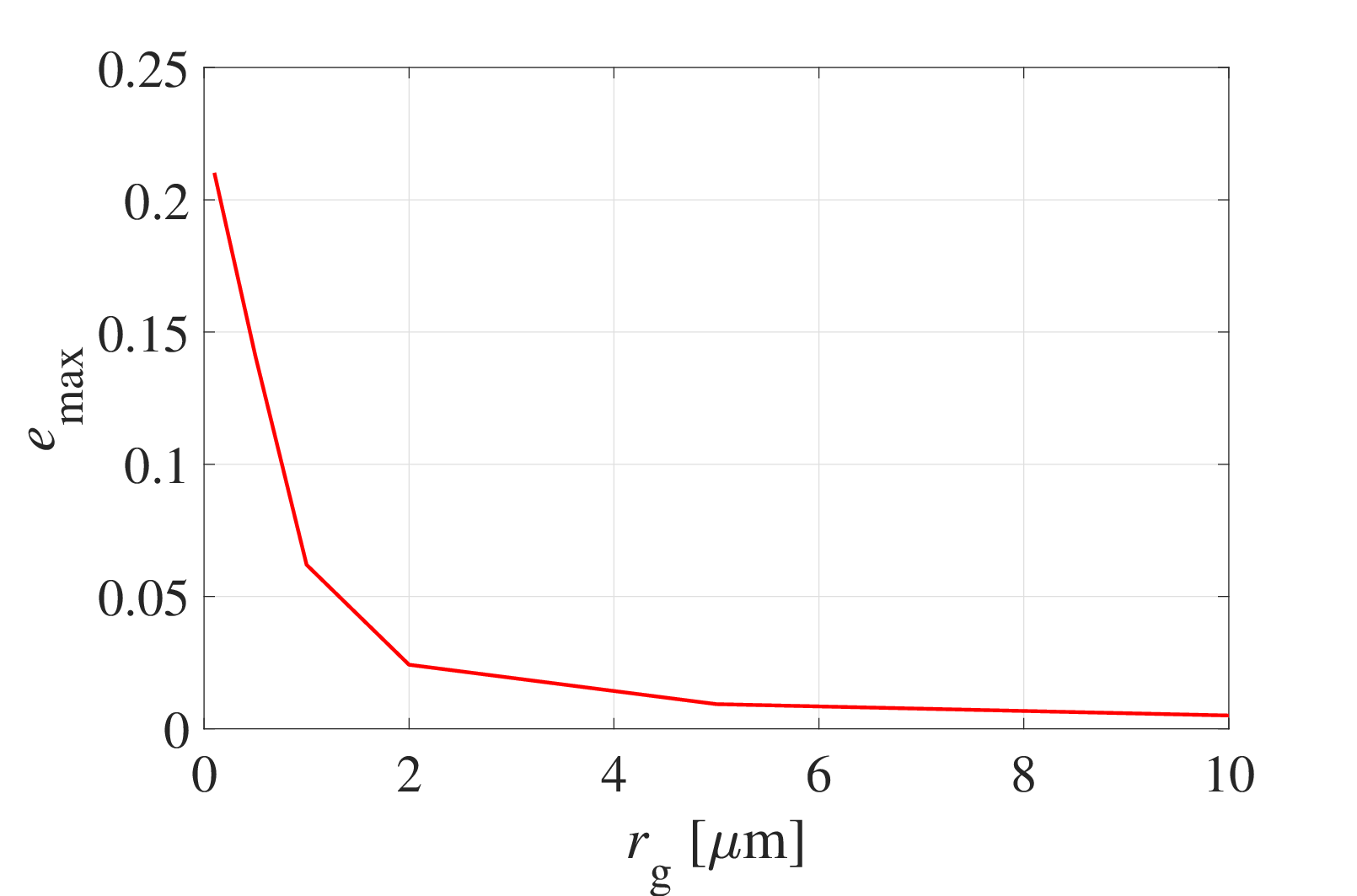}
    \caption{%The 
    Variation in the maximum value of the eccentricity of a particle with initial size of $10\,{\mu}$m as its grain size decreases, which is calculated from the conserved integral of dust particles' motion derived by \citet{hamilton1996circumplanetary}.}
    \label{emax}
\end{figure}

The conserved integral of dust particles' motion derived by \citet{hamilton1996circumplanetary} is used to calculate the variation in the maximum value of the eccentricity $e_\mathrm{max}$ of a particle with initial size of $10\,{\mu}$m as its grain size decreases (see Fig.~\ref{emax}).
%The $e_\mathrm{max}$ increases slowly and remains small during the grain size between $2\,{\mu}$m to $10\,{\mu}$m.
%After the grain size smaller than $2\,{\mu}$m, the $e_\mathrm{max}$ begins to increase quickly as the grain size decreases.
The value of $e_\mathrm{max}$ remains small in the grain size of $[2,\,10]~\mathrm{\mu}$m.
After the grain size is smaller than $2\,{\mu}$m, the value of $e_\mathrm{max}$ increases quickly as the grain size decreases.
This analytical solution matches well with our numerical results.
Note that the value of $e_\mathrm{max}$ calculated by the conserved integral may not be exact when the particle is smaller than $2\,{\mu}$m since the orbit-average approximation for deriving the integral breaks down in this size range. However, the basic variation trend of $e_\mathrm{max}$ is still reliable \citep{hamilton1996circumplanetary}.

\begin{figure}
        \includegraphics[width=\columnwidth]{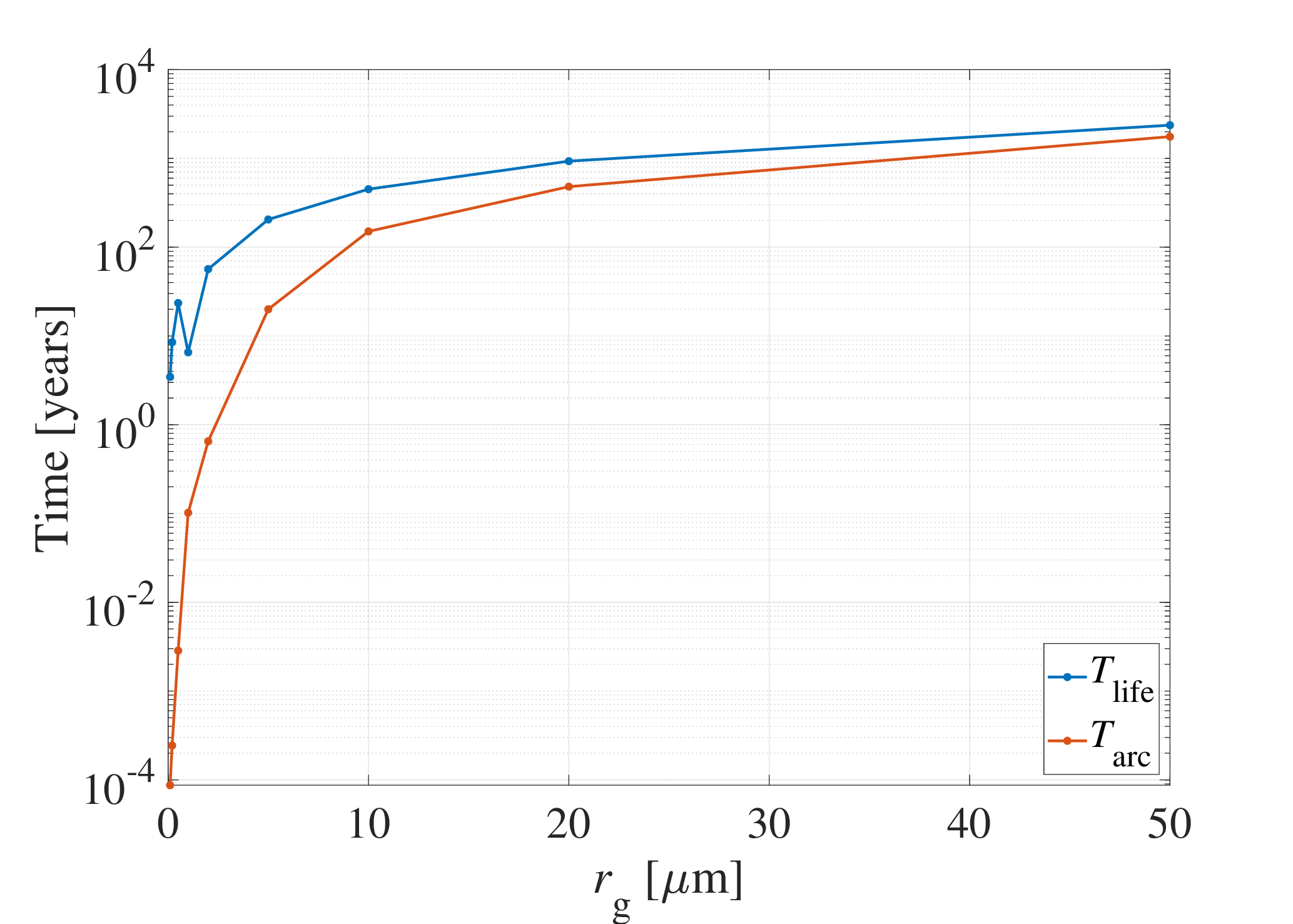}
    \caption{Average lifetime $T_\mathrm{life}$ and average time $T_\mathrm{arc}$ that dust particles spend in the arc as a function of grain size.}
    \label{lifetime}
\end{figure}

Based on the results of the dynamical evolution of dust particles, we analyze the average dynamical lifetimes $T_\mathrm{life}$, the average times $T_\mathrm{arc}$ that dust particles spend in the arc, and the final fates of particles with different sizes (see Fig.~\ref{lifetime} and Table.\ref{final}).
Note that $T_\mathrm{arc}$ is defined to be the duration from the production moment of particle to the escaping moment of particle from the arc.
\begin{table}
	%\centering
	\caption{%The 
 Fractions of different final fates of dust particles with different grain sizes. Note that the 'Move outside' column represents the fraction of particles that migrate outside the orbit of Enceladus. The fractions of dust particles that hit Enceladus is very small and not listed here.}
	%Remember to define the quantities, symbols and units used.}
        \resizebox{\columnwidth}{!}{
        \label{final}
        \renewcommand{\arraystretch}{1.5}
        \fontsize{25pt}{25pt}\selectfont
	\begin{tabular}{lcccccr} % four columns, alignment for each
		\hline
		$r_\mathrm{g}(\mathrm{\mu m})$ & Removed by sputtering & Hit A ring & Move outside & Hit Mimas & Hit Aegaeon\\
		\hline
		0.1 & 0.988 & 0 & 0 & 0 & 0.012\\
		0.2 & 1.0 & 0 & 0 & 0 & 0\\
		0.5 & 1.0 & 0 & 0 & 0 & 0\\
            1.0 & 0 & 0.955 & 0 & 0.023 & 0.022\\
            2.0 & 0 & 0.947 & 0 & 0.035 & 0.018\\
            5.0 & 0 & 0.899 & 0 & 0.099 & 0.002\\
            10.0 & 0 & 0.473 & 0.315 & 0.197 & 0.003\\
            20.0 & 0 & 0.414 & 0.252 & 0.33 & 0.003\\
            50.0 & 0 & 0.478 & 0.148 & 0.353 & 0.019\\
		\hline
	\end{tabular}}
\end{table}
The $T_\mathrm{arc}$ of dust particles smaller than $1\,{\mu}$m are less than one year, and the $T_\mathrm{life}$ of these particles are determined by the sputtering rate as they are finally removed by the plasma sputtering.
Most of $1\,{\mu}$m particles collide with the A ring in a few years due to their orbital distances oscillating with large amplitude. 
For particles larger than $1\,{\mu}$m, the $T_\mathrm{arc}$ increases with the initial grain size since the decrease in grain size due to the plasma sputtering before the particles leave the arc is proportional to the initial grain size (note that the sputtering rate is the same for particles regardless of their sizes).
The dynamical lifetimes of particles after they leave the arc are mainly determined by the time span when the eccentricities of particles remain small (see Fig.~\ref{orbital evolution}c), and also increase with the initial grain size. By adding up the dynamical lifetime of particles before and after leaving the arc, the $T_\mathrm{life}$ also increase with the initial grain size.
As seen from Table.~\ref{final}, most of dust particles with $r_\mathrm{g}>1\,{\mu}$m collide with the A ring due to the close proximity between the orbital distance of the outer edge of the A ring ($\sim$$2.27\,R_\mathrm{s}$) and the initial orbital distances of particles ($\sim$$2.75\,R_\mathrm{s}$) and the significant oscillation of the orbital distances of particles.
However, a fraction of large particles ($r_\mathrm{g}\geq10\,{\mu}$m) can also migrate outside the orbit of Enceladus ($\sim$$3.95\,R_\mathrm{s}$) due to their long dynamical lifetimes after they leave the arc, which allows their semimajor axes (approximately equal to their orbital distances when the eccentricities are small) to increase to that of Enceladus before their orbital distances oscillating significantly.
During the migration, a fraction of particles hit Mimas ($\sim$$3.09\,R_\mathrm{s}$).
As large particles move outward slower than small particles and have a larger probability of hitting Mimas, the fraction of particles that hit Mimas increases with the grain size.
%A fraction of particles hit Mimas ($\sim$$3.09\,R_\mathrm{s}$), \textcolor{red}{and the fraction increases with the grain size as large particles move outward slower than small particles due to the plasma drag and have a larger probability of hitting Mimas.}
We note that the fraction of particles that hit Aegaeon is very small. This may be attributed to various non-gravitational effects considered in our simulations due to small grain sizes, which makes the distribution of particles dispersed.

\section{Properties of the G ring}
\label{Properties of the G ring}
Based on the simulation results of the dynamical evolution of dust particles originating from the G-ring arc, the properties of the simulated ring formed by these particles, including the normal $I/F$, the number density of dust particles and the geometric optical depth, can be estimated by integrating the simulated trajectories of dust particles over their size distribution, which reads \citep{brooks2004size,liu2021configuration}
\begin{equation}
\resizebox{\columnwidth}{!}{$
    4\mu\frac{I}{F}(i_\mathrm{cell},j_\mathrm{cell})=\int_{r_\mathrm{min}}^{r_\mathrm{max}}\frac{\tilde{n}(i_\mathrm{cell},j_\mathrm{cell}; r_\mathrm{g}){\pi}r_\mathrm{g}^2Q_\mathrm{sca}(r_\mathrm{g})P(r_\mathrm{g},\alpha)}{S_\mathrm{normal}(i_\mathrm{cell},j_\mathrm{cell})n_\mathrm{start}(r_\mathrm{g})}{\Delta}t N(r_\mathrm{g})\mathrm{d}r_\mathrm{g},
    $}
    \label{eq:normalif}
\end{equation}
\begin{equation}
    \resizebox{8cm}{!}{$
        \begin{split}        
\tau(i_\mathrm{cell},k_\mathrm{cell})=\int_{r_\mathrm{min}}^{r_\mathrm{max}}\frac{\tilde{n}(i_\mathrm{cell},k_\mathrm{cell}; r_\mathrm{g}){\pi}r_\mathrm{g}^2}{S_\mathrm{edge-on}(i_\mathrm{cell},k_\mathrm{cell})n_\mathrm{start}(r_\mathrm{g})}{\Delta}t N(r_\mathrm{g})\mathrm{d}r_\mathrm{g},
        \end{split}
    $}
    \label{eq:optical}
\end{equation}
\begin{equation}
\resizebox{8cm}{!}{$
n(i_\mathrm{cell},j_\mathrm{cell},k_\mathrm{cell})=
\int_{r_\mathrm{min}}^{r_\mathrm{max}}\frac{\tilde{n}(i_\mathrm{cell},j_\mathrm{cell},k_\mathrm{cell}; r_\mathrm{g})}{V(i_\mathrm{cell},j_\mathrm{cell},k_\mathrm{cell})n_\mathrm{start}(r_\mathrm{g})}{\Delta}t N(r_\mathrm{g})\mathrm{d}r_\mathrm{g}$},
     \label{eq:numberdensity}
\end{equation}
%The normal $I/F$ in the region with a surface area of $S_\mathrm{normal}(i_\mathrm{cell},j_\mathrm{cell})$ when viewed from above is represented by $\mu\frac{I}{F}(i_\mathrm{cell},j_\mathrm{cell})$.
where the space around the ring is divided into a series of grid cells indexed by $(i_\mathrm{cell},j_\mathrm{cell},k_\mathrm{cell})$ in the Saturn-centered cylindrical coordinate system. The variable $\mu\frac{I}{F}(i_\mathrm{cell},j_\mathrm{cell})$ denotes the normal $I/F$ in the region with a surface area of $S_\mathrm{normal}(i_\mathrm{cell},j_\mathrm{cell})$ when viewed from above, $r_\mathrm{min}$ and $r_\mathrm{max}$ are the minimum and maximum grain sizes of dust particles in our simulations, $\tilde{n}(i_\mathrm{cell},j_\mathrm{cell}; r_\mathrm{g})$ is the vertically-integrated number of times that dust grains with grain size $r_\mathrm{g}$ pass the region indicated by $(i_\mathrm{cell},j_\mathrm{cell})$, $Q_\mathrm{sca}(r_\mathrm{g})$ and $P(r_\mathrm{g},\alpha)$ are the scattering efficiency and the phase function of dust particles, and both can be calculated by Mie theory \citep{brooks2004size,throop2004jovian}, $n_\mathrm{start}(r_\mathrm{g})$ is the number of simulated particles for each grain size, ${\Delta}t$ is the constant time-step used for segmenting the trajectories of particles, $N(r_\mathrm{g})$ is the particle size distribution, $\tau(i_\mathrm{cell},k_\mathrm{cell})$ is the edge-on geometric optical depth of the ring, $\tilde{n}(i_\mathrm{cell},k_\mathrm{cell}; r_\mathrm{g})$ is the number of times that dust particles with grain size $r_\mathrm{g}$ pass the region with a surface area of $S_\mathrm{edge-on}(i_\mathrm{cell},k_\mathrm{cell})$ when viewed edge-on, $n(i_\mathrm{cell},j_\mathrm{cell},k_\mathrm{cell})$ is the number density of dust particles in the grid cell with volume $V(i_\mathrm{cell},j_\mathrm{cell},k_\mathrm{cell})$, $\tilde{n}(i_\mathrm{cell},j_\mathrm{cell},k_\mathrm{cell}; r_\mathrm{g})$ is the number of times that dust particles with grain size $r_\mathrm{g}$ pass the grid cell ($i_\mathrm{cell},j_\mathrm{cell},k_\mathrm{cell}$).

The detailed study of the size distribution of dust particles in the G ring shows that the size of dust particles within the G ring can be characterized by a power-law distribution with the exponent $q$ in the range of [1.5, 3.5] \citep{throop1998g}. 
Results from other studies also support that the size distribution of particles inside the G ring is relatively flat and the value of $q$ is smaller than 3.5 \citep{meyer1998constraints,de2004keck}.
Here we assume that the particle size distribution of the ring follows the differential power-low distribution, which is expressed as
\begin{equation}
    N(r_\mathrm{g})\mathrm{d}r_\mathrm{g}=C_0r_\mathrm{g}^{-q}\mathrm{d}r_\mathrm{g},
    \label{eq:sizedistribution}
\end{equation}
where $C_0$ is the normalization constant.
%\textcolor{red}{, which was} not strongly constrained in previous studies.
%Since there are two unknowns, namely $C_0$ and $q$ 
%we derive the best solution of particle size distribution by making both the normal $I/F$ estimated from Eq.~(\ref{eq:normalif}) and the number density estimated from Eq.~(\ref{eq:numberdensity}) match well with that inferred from the observations \citep{hedman2007source,ye2016situ}.
%Finally, we get the $q=2.73$ and $C_0=12.7$.
By matching both the normal $I/F$ estimated from Eq.~(\ref{eq:normalif}) and the number density estimated from Eq.~(\ref{eq:numberdensity}) with that inferred from the observations \citep{hedman2007source,ye2016situ}, we get $q=2.8$ and $C_0=4.0$.

\begin{figure}
        \includegraphics[width=\columnwidth]{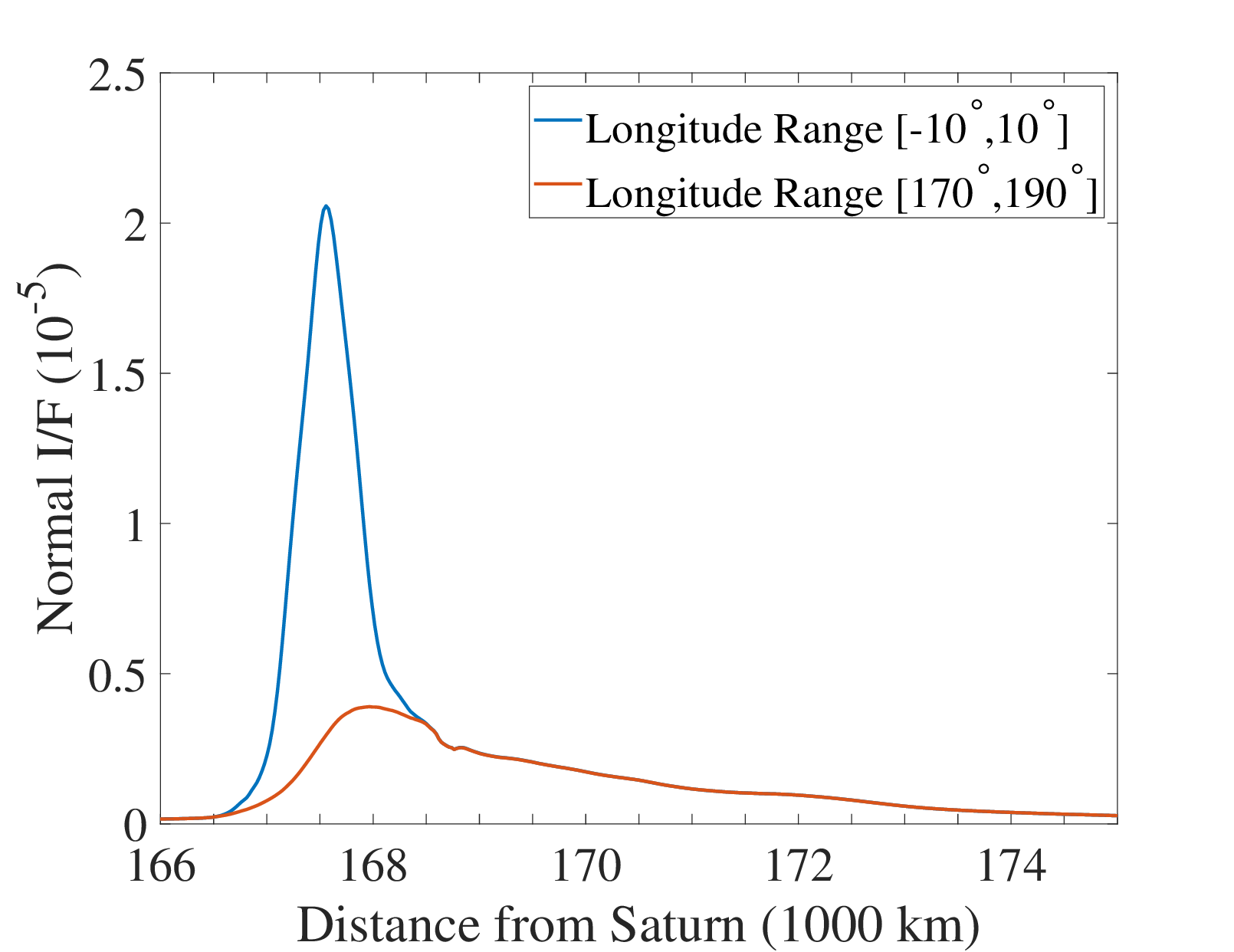}
    \caption{Radial profiles of the normal $I/F$ of the simulated ring formed by dust particles originating from the G-ring arc, which is derived from Eq.~(\ref{eq:normalif}) with $q=2.8$ and $C_0=4.0$. The profile in the longitude range of $[-10^\circ, 10^\circ]$ passes through the arc, and the profile in the longitude range of $[170^\circ, 190^\circ]$ passes through elsewhere. The peak value of the normal $I/F$ in the longitude range of $[170^\circ, 190^\circ]$ is aligned with that derived from the observation \citep[see][Fig.~2]{hedman2007source}.}
    \label{figIF}
\end{figure}

The average radial normal $I/F$ profiles derived from Eq.~(\ref{eq:normalif}) %with $q=2.73$ and $C_0=12.7$ 
are shown in Fig.~\ref{figIF}, where the profile in the longitude range of $[-10^\circ, 10^\circ]$ passes through the arc, and the profile in the longitude range of $[170^\circ, 190^\circ]$ passes through elsewhere.
One of the fits for deriving the $q=2.8$ and $C_0=4.0$ is that aligns the peak value of the simulated normal $I/F$ in the longitude range of $[170^\circ, 190^\circ]$ with that of the G ring, which is obtained from the observation \citep[see][Fig.~2]{hedman2007source}.
With $q=2.8$ and $C_0=4.0$, the normal $I/F$ profiles from our numerical simulations match well with that derived by \citet{hedman2007source}, which supports that the G ring may originate from the arc.
The radial range of the ring is $[166,000, 175,000]\,\mathrm{km}$.
%In the part of the ring that does not pass through the arc
For the profile through elsewhere, the normal $I/F$ increases quickly to the peak value of $3.9\times10^{-6}$ at the radial distance of $\sim$$168,000\,\mathrm{km}$, and then decreases slowly with the increase of the radial distance from Saturn.
%In the part of the ring which passes through the arc,
For the profile through the arc, the normal $I/F$ increases and decreases sharply within the radial range of $2,000\,\mathrm{km}$.
Subsequently, the normal $I/F$ decreases slowly, following the similar trend of the profile through elsewhere.
Therefore, most of the arc is confined in the radial range of $[166,500, 168,500]\,\mathrm{km}$.
The peak position of the normal $I/F$ of the arc is about $167,500\,\mathrm{km}$, which is close to the 7:6 CER with Mimas \citep{hedman2007source}.

\begin{figure}
        \includegraphics[width=\columnwidth]{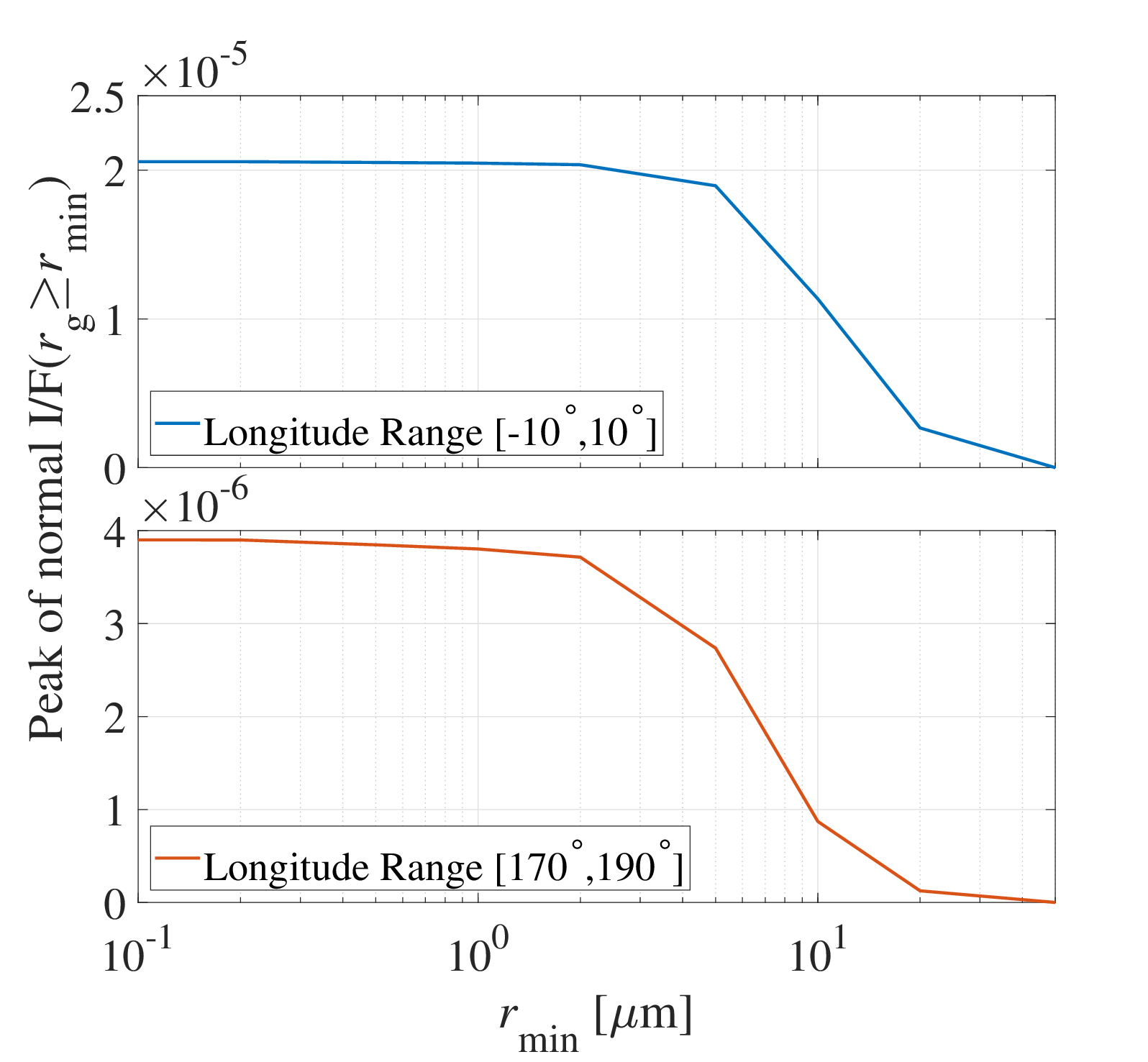}
    \caption{Cumulative distributions of the peak of the normal $I/F$ of different radial normal $I/F$ profiles of the simulated ring in Fig.~\ref{figIF}.} 
    % where the part of the ring in the longitudinally range of $[-10^\circ, 10^\circ]$ passes through the arc, and the part of the ring in the longitudinally range of $[170^\circ, 190^\circ]$ do not pass through the arc.}
    \label{figIFcontri}
\end{figure}

We note that the average normal $I/F$ in the longitude range of $[-10^\circ, 10^\circ]$ of our results is slightly higher than that derived by \citet{hedman2007source}, the peak values of which are $2.1\times10^{-5}$ and $1.6\times10^{-5}$, respectively.
%Since the difference only appears in the profile through the arc, the difference between different parts of the ring (passes through the arc or not) is investigated.
To analyze the dynamical cause for this difference, we calculate the cumulative distributions of the peak of the normal $I/F$ of different radial normal $I/F$ profiles, as shown in Fig.~\ref{figIFcontri}.
The normal $I/F$ in the longitude range of $[-10^\circ, 10^\circ]$ is mainly contributed by dust particles in the range of $[5, 20]\,{\mu}$m, followed by the contribution from dust particles larger than $20\,{\mu}$m.
%and dust particles larger than $20\,{\mu}$m also make contributions to the normal $I/F$ of the arc.
%In the part of the ring that does not pass the arc
For the profile through elsewhere, the contributions of $[10, 50]\,{\mu}$m dust particles to the normal $I/F$ are less than their contributions to the normal $I/F$ profile through the arc.
For these large particles ($r_\mathrm{g}\geq10\,\mathrm{{\mu}m}$), they stay in the arc longer than the small particles (see Fig.~\ref{lifetime}), and therefore have a higher probability of colliding with the parent bodies in the arc. However, these collisions are not involved in our simulations due to the difficulties of simulating the orbits of a large number of parent bodies.
%\citet{madeira2018production} \textcolor{cyan}{shows that 75\% of particles confined in the G arc ultimately collide with Aegaeon within 500 years.}
%if dust particles are confined in the G arc, 75 percent of them collides with Aegaeon within 500 years. 
%Note that the collision of dust particles with Aegaeon is difficult to be detected in our numerical simulation \textcolor{cyan}{due to the small size of Aegaeon and the simplification of its orbit into a precessing ellipse.}
%, which is because Aegaeon is too small and its orbit is simply assumed to be a precessing ellipse. 
%Based on the result of \citet{madeira2018production}, we believe that if the collision of particles with other large bodies in the arc is also considered, the fraction and frequency of the collision must be higher.
Therefore, we suggest that the slightly higher value of the normal $I/F$ in the longitude range of $[-10^\circ, 10^\circ]$ of our results compared to that derived by \citet{hedman2007source} comes from the ignorance of these collisions.
%that we do not consider the removal of dust particles due to the collision with the parent bodies in the arc.

\begin{figure}
        \includegraphics[width=\columnwidth]{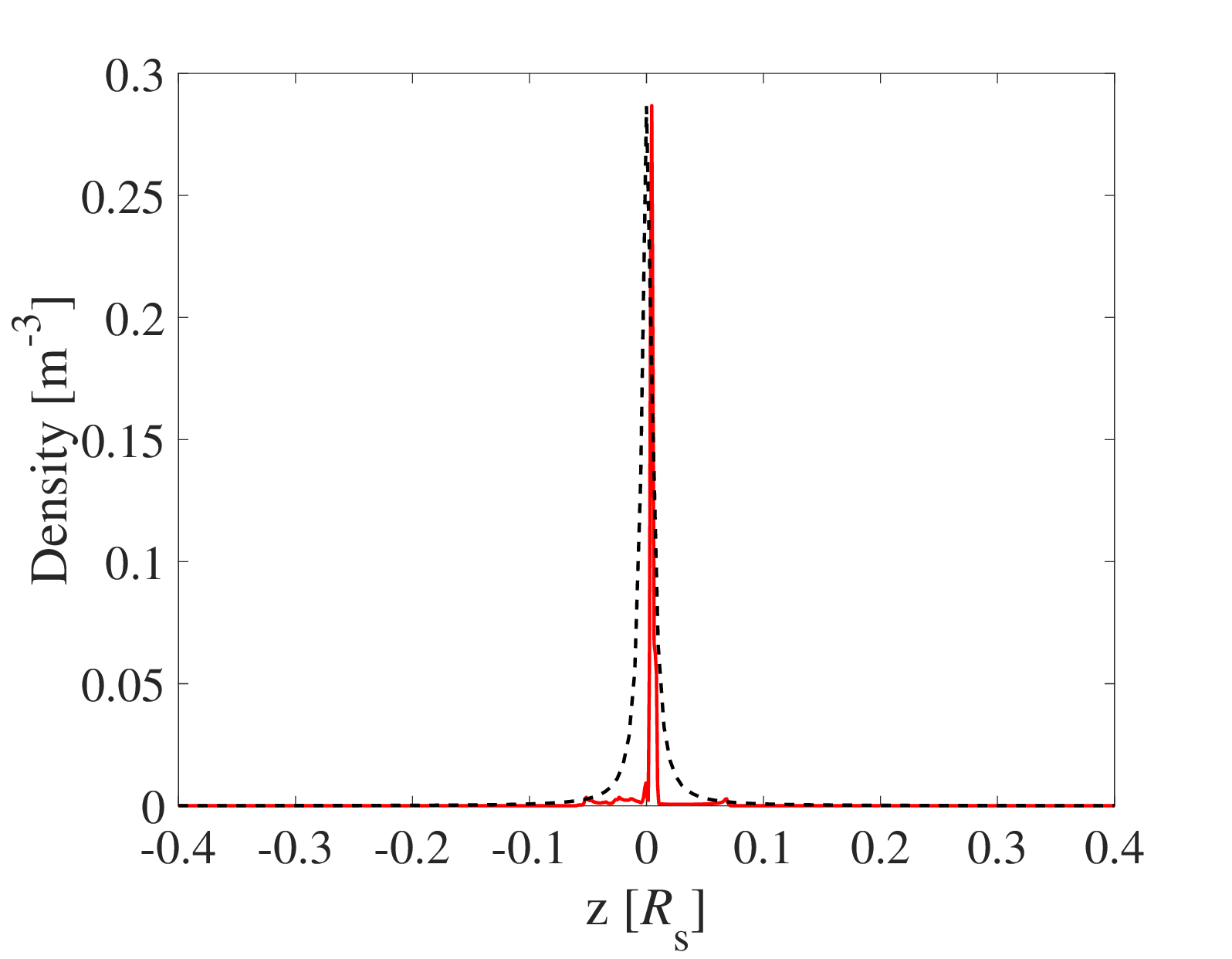}
    \caption{Vertical number density profiles at the radial distance of $2.7\,R_\mathrm{s}$. The red line and the black dash line denote the simulated profiles derived from Eq.~(\ref{eq:numberdensity}) with $q=2.8$ and $C_0=4.0$ and from the in-situ measurement by the Cassini spacecraft \citep{ye2016situ}, respectively.}
    \label{fignemberdensity}
\end{figure}

\begin{figure}
        \includegraphics[width=\columnwidth]{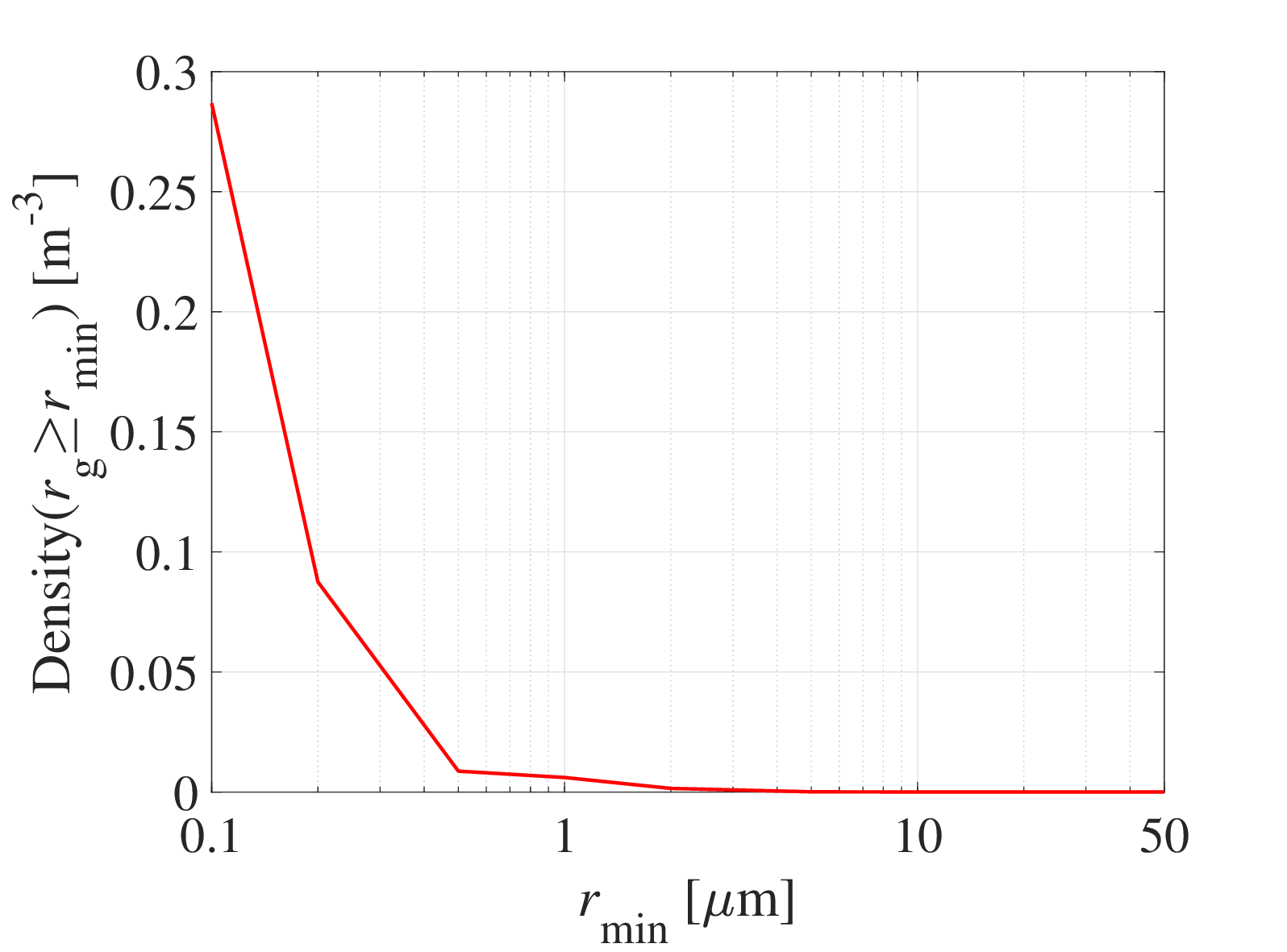}
    \caption{Cumulative distribution of the peak vertical number density in Fig.~\ref{fignemberdensity}.}
    \label{fignemberccontri}
\end{figure}

Another fit to get the $q=2.8$ and $C_0=4.0$ is that aligns the peak value of the vertical number density at the radial distance of $2.7\,R_\mathrm{s}$ with that derived by \citet{ye2016situ}, as shown in Fig.\,\ref{fignemberdensity}.
The red line is the vertical number density profile estimated from Eq.~(\ref{eq:numberdensity}), and the black dash line is the fitting result derived from the in-situ measurement by the Cassini spacecraft \citep{ye2016situ}. 
The vertical range of dust particles in both results is about $[-0.1, 0.1]\,R_\mathrm{s}$.
The position of the peak number density is $0.005\,R_\mathrm{s}$ above the equatorial plane in our result, while that derived from the in-situ measurement is close to the equatorial plane.
%Additionally, the vertical number density from our result is much lower than that derived from the in-situ measurement in most of the vertical range.
%the vertical number density in our result increases and decreases sharply in a \textcolor{cyan}{narrow} vertical range.
%\textcolor{cyan}{Except for the narrow range near the peak position, the vertical number density from our result is much lower than that derived from the in-situ measurement in most of the vertical range.}
The difference between our result and that derived by \citet{ye2016situ} may come from the assumption that the electric potential of dust particles is a constant (see Section.~\ref{Dynamical model}).
For dust particles smaller than $0.5\,{\mu}$m, which contribute the most to the observed number density (see Fig.~\ref{fignemberccontri}), this assumption may minimize the variation of the inclination of particles due to the Lorentz force related to the electric potential of particles, which results in the shift of the position of peak vertical number density and the concentrated distribution of dust particles in the vertical direction.
%From the cumulative distribution of the peak number density of particles (shown in Fig.~\ref{eq:numberdensity}), we find that dust particles smaller than $0.5\,{\mu}$m contribute the most to the number density.
%These small particles are strongly affected by the Lorentz force which is related to the dust particle's potential.}
%To \textcolor{cyan}{analyze} the dynamical reason for these differences, we first find out the grain sizes of particles which contribute significantly to the number density by calculating the cumulative distribution of the peak number density (see Fig.~\ref{fignemberccontri}).
%Dust particles smaller than $0.5\,{\mu}$m are found to be the most important.
% From the cumulative distribution of the peak number density of particles (shown in Fig.~8), we find that dust particles small than $0.5\,{\mu}$m contribute the most to the number density.  
%For these small particles, the Lorentz force related to the dust particle's potential is important\textcolor{blue}{significant}.
%Therefore, the differences between the vertical number density of our result and that derived by \citet{ye2016situ} may come from the assumption that the potential of dust particle is a constant (see Sect.~\ref{Dynamical model}).


\begin{figure}
        \includegraphics[width=\columnwidth]{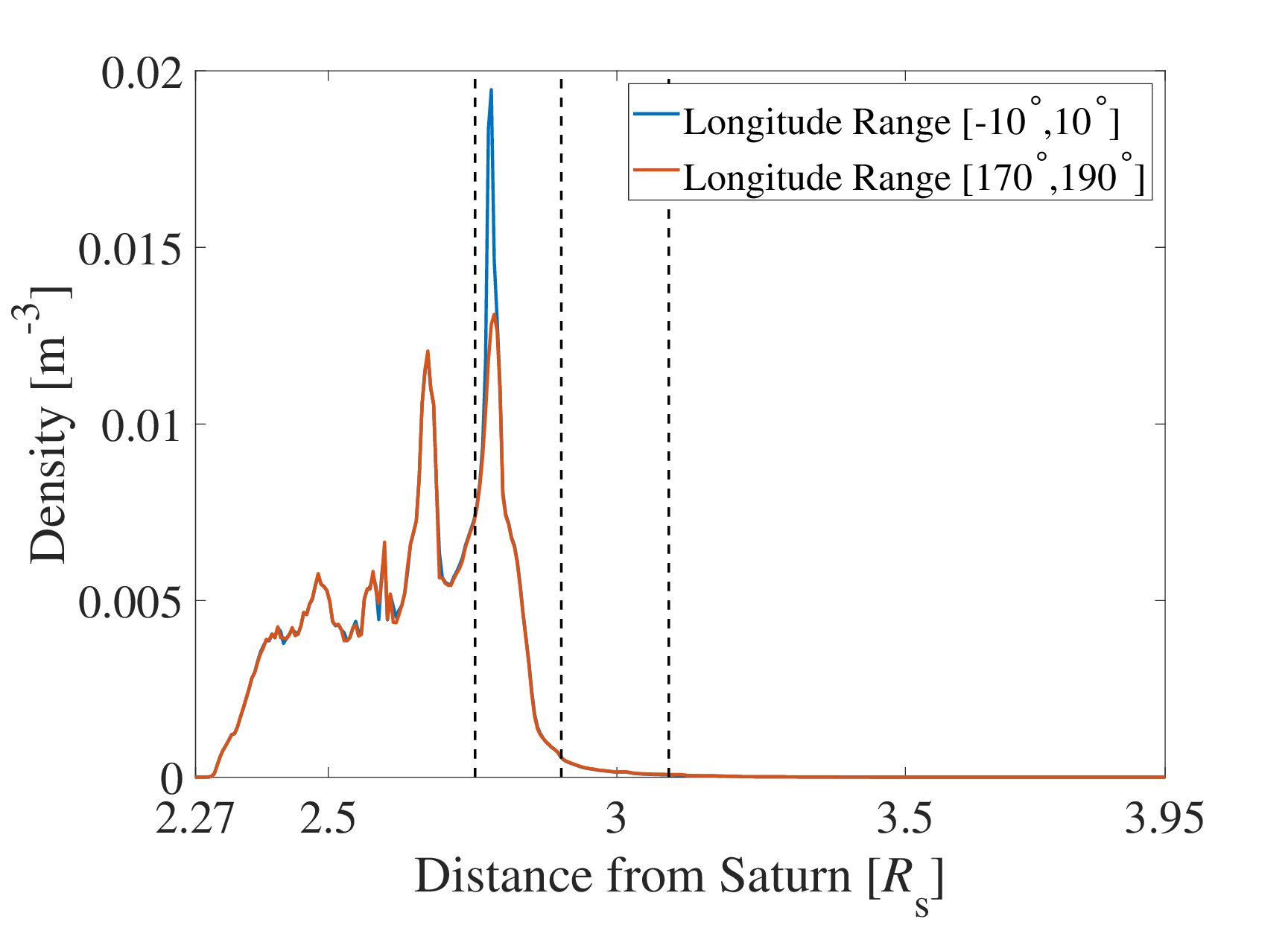}
    \caption{Simulated radial number density profiles (averaged over the vertical range of $[-0.1, 0.1]\,R_\mathrm{s}$) of dust particles, where the profile in the longitude range of $[-10^\circ, 10^\circ]$ passes through the arc, and the profile in the longitude range of $[170^\circ, 190^\circ]$ passes through elsewhere. The three black dash lines from left to right represent the orbital distances of the inner edge of the G ring $(166,000\,\mathrm{km}\approx2.75\,R_\mathrm{s})$, the outer edge of the G ring $(167,500\,\mathrm{km}\approx2.78\,R_\mathrm{s})$ and Mimas $(186,000\,\mathrm{km}\approx3.09\,R_\mathrm{s})$, respectively.}
    \label{number_r}
\end{figure}

\begin{figure}
        \includegraphics[width=\columnwidth]{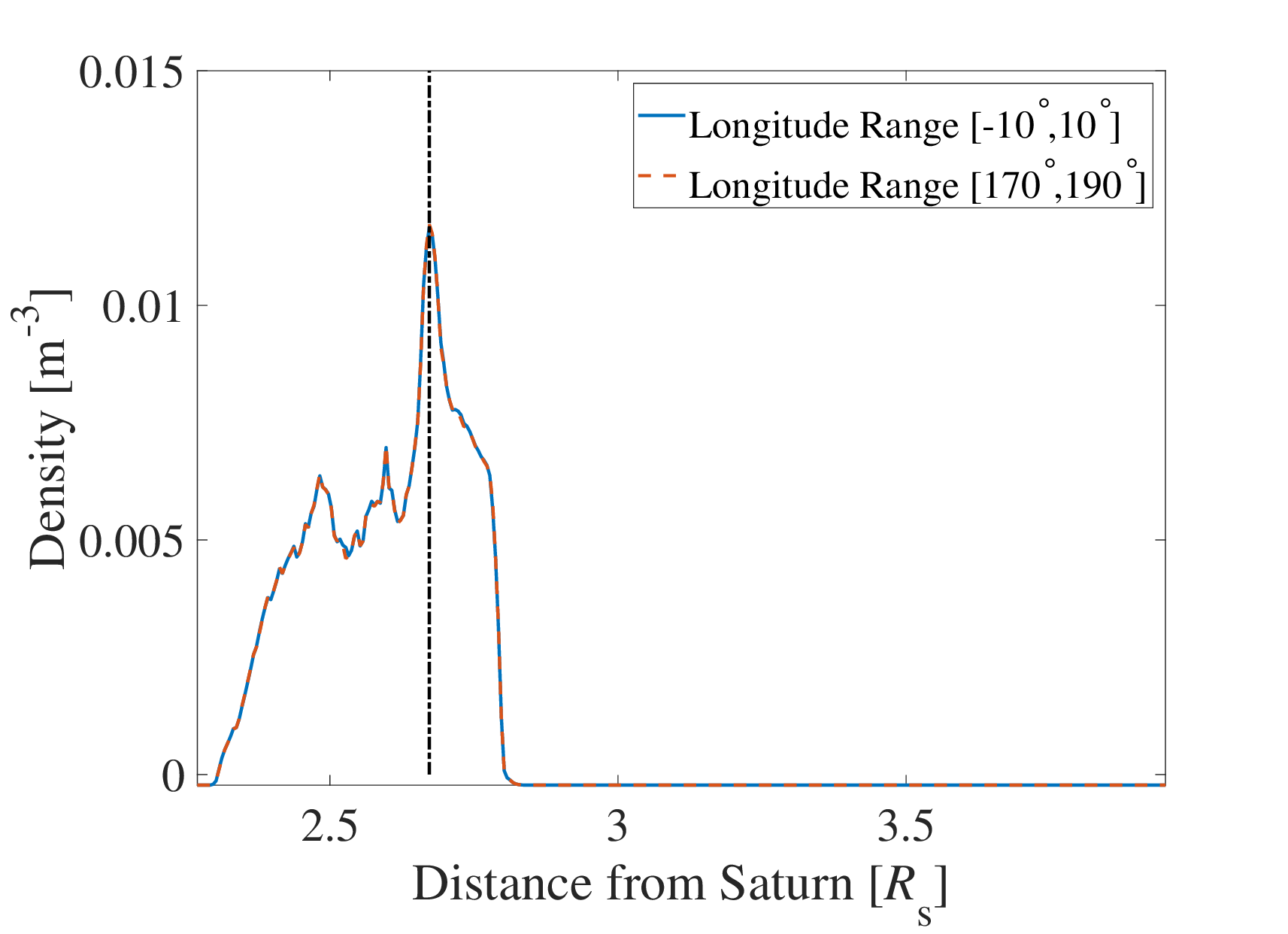}
    \caption{Simulated radial number density profiles of particles smaller than $0.5\,{\mu}$m (averaged over the vertical range of $[-0.1, 0.1]\,R_\mathrm{s}$), where the profile in the longitude range of $[-10^\circ, 10^\circ]$ passes through the arc, and the profile in the longitude range of $[170^\circ, 190^\circ]$ passes through elsewhere. The black dot-dashed line represents the orbital distance of 2.67$\,R_\mathrm{s}$.}
    \label{number_small}
\end{figure}

%With $q=2.73$ and $C_0=12.7$, 
The average radial number density profiles (averaged over the vertical range of $[-0.1, 0.1]\,R_\mathrm{s}$) of dust particles originating from the G-ring arc in different longitude ranges are also derived from Eq.~(\ref{eq:numberdensity}), as plotted in Fig.~\ref{number_r}.
%The profile in the longitude in the range of $[-10^\circ, 10^\circ]$ passes through the arc, and the profile in the longitude in the range of $[170^\circ, 190^\circ]$ passes through elsewhere.
The number density profiles of particles in these two longitude ranges are largely similar, except for the region of the arc with a peak radial number density of about $0.019\,\mathrm{m}^{-3}$.
In addition to the G ring, a lot of dust particles originating from the G-ring arc are also distributed in the region between the A ring and the G ring, and there is another peak position of the number density around $2.67\,R_\mathrm{s}$.
%\textcolor{red}{As shown in Fig.~\ref{number_small}, particles in this region and the peak at $2.67\,R_\mathrm{s}$ are attributed to particles smaller than $0.5\,{\mu}$m.}
This peak at $2.67\,R_\mathrm{s}$ may be attributed to the fact that particles smaller than $0.5\,{\mu}$m dominate in this region (see Fig.~\ref{fignemberccontri}), and their number density peaks around this radial distance (see Fig.~\ref{number_small}).
%spend considerable time around this radial distance.
%, which is related to the scattering efficiency and the phase function of particles smaller than $0.5\,{\mu}$m that are dominant in this region (see Fig.~\ref{fignemberccontri}).}
%Although the peak value of the number density in $2.67\,R_\mathrm{s}$ is close to that of the ring (the part that do not pass the arc), there is no a extension of the G ring has been found here.
%This is because the dominant grain sizes of particles are different in the region inside the ring and in the region of the ring, as shown in Fig.~\ref{figIFcontri} and Fig.~\ref{fignemberccontri}.
In the region outside the G ring, the number density is small and decreases to nearly zero outside the orbital distance of Mimas.

The edge-on geometric optical depth of the simulated ring is calculated from Eq.~(\ref{eq:optical}).
From this result, the apparent edge-on thickness of the G ring which has not been resolved from observations \citep{nicholson1996observations} can be constrained (see Fig.~\ref{optical depth}, where the line of sight passes through the arc).
%by setting different lower limit of the edge-on geometric optical depth that can be observed by an observing instrument (see Fig.~\ref{optical depth}, where the line of sight passes through the arc).
If the observable threshold of the edge-on geometric optical depth of an observing instrument is $1\times10^{-8}$, the apparent edge-on thicknesses of the inner and outer edges of the G ring are about $9,000\,\mathrm{km}$ and $3,000\,\mathrm{km}$, respectively.
%If the lower limit of the edge-on geometric optical depth that can be observed by an observing instrument is $1\times10^{-8}$, the apparent edge-on thicknesses of the inner and outer edges of the G ring are about $9,000\,\mathrm{km}$ and $3,000\,\mathrm{km}$, respectively.
If the observable threshold is a higher value of $1\times10^{-6}$, the ring is only observable within a north-south thickness of $650\,\mathrm{km}$, and the vertical range of the ring in the north of Saturn's equatorial plane is tens of kilometers larger than that in the south.
%In fact, most of the ring is confined to a vertical range of $50\,\mathrm{km}$ from the equatorial plane of Saturn, with the average edge-on geometric optical depth of the order of $10^{-4}$.
In the region of the arc, the edge-on geometric optical depth peaks at a value of $3.9\times10^{-2}$.
Note that when the line of sight does not pass through the arc, the apparent edge-on thickness of the G ring is similar, but the maximum value of the edge-on geometric optical depth is on the smaller order of $10^{-3}$.
\begin{figure}
        \includegraphics[width=\columnwidth]{optical_depth_1e-8.eps}
        \includegraphics[width=\columnwidth]{optical_depth_1e-6.eps}
    \caption{\textit{Panel a:} edge-on geometric optical depth of the simulated ring, where the line of sight passes through the arc; only the values of edge-on geometric optical depth larger than $1\times10^{-8}$ are shown. \textit{Panel b:} same as \textit{Panel a}, but only the values of edge-on geometric optical depth larger than $1\times10^{-6}$ are shown.}
    %Panel (a) shows the part of the ring with optical depth larger than $1\times10^{-8}$, and panel (b) shows the part of the ring with optical depth larger than $1\times10^{-6}$.}}
    \label{optical depth}
\end{figure}
%Crude estimates of the age and the remaining lifetime of the G ring can be computed based on the result of the particle size distribution of the ring.

%Since the particle size distribution of the arc and the ring is assumed to be consistent (see section 5), i.e., the size distribution do not change after dust particles escaping from the arc to the ring, 

The mass of dust particles escaping from the arc to the ring per second $M^+$ can be calculated by integrating the particle size distribution, which reads
\begin{equation}
    M^+ = \int_{r_\mathrm{min}}^{r_\mathrm{max}}\frac{4}{3}{\pi} \, r_\mathrm{g}^{3} \,{\rho_\mathrm{g}} \, N(r_\mathrm{g}) \, \mathrm{d}r_\mathrm{g}.
        \label{eq:production}
\end{equation}
By substituting Eq.~(\ref{eq:sizedistribution}), we %can 
get $M^+ =5.4\times10^{-2}\,\mathrm{kg\,s^{-1}}$.
%\textcolor{cyan}{, which is about three orders larger than the mass flux ejected from Aegaeon ($\sim$ $10^{-5}\,\mathrm{kg\,s^{-1}}$,  estimated by Muñoz-Gutierrez et al. 2021)}.
The mass of the arc has been estimated by \citet{hedman2007source}, where the value is $10^8$-$10^{10}\,\mathrm{kg}$.
Assuming that the $M^+$ is constant and the G ring is only maintained by the G-ring arc, the upper limit of the remaining lifetime of the ring can be estimated to be %in
on the order of $10^{4}\,\mathrm{years}$ by dividing the total mass of the parent bodies by the $M^+$. This is a threshold value since the entire mass of arc may not be fully converted to the ring.
The mass production rate of dust particles from Aegaeon was estimated by \citet{madeira2018production} and \citet{munoz2022long} to be on the order of $10^{-7}\,\mathrm{kg\,s^{-1}}$ and $10^{-5}\,\mathrm{kg\,s^{-1}}$, respectively, which is much smaller than the $M^+$. This suggests that the G ring is dominated by the dust particles from the arc.
% Therefore, we suggest that other sources of dust particles in the arc, such as the collisions between large debris, should produce dust at a rate comparable to $M^+$ to maintain the steady state of the arc and the ring.
%\textcolor{red}{Otherwise, the remaining lifetime of the ring is very short if it is only maintained by dust particles already exist in the arc, the mass of which is about $2\times10^{6}\,\mathrm{kg}$.} This is consistent with the conclusions of \citet{madeira2018production}, which suggested that Aegaeon is not the only source of particles within the arc.
%The dust particles already existing in the arc would run out and the G ring would disappear within several years if Aegaeon is the only source. Nevertheless, such a rapid disappearance of the ring is unlikely. Therefore, we suggest that Aegaeon is not the only source of particles within the arc, which is consistent with the conclusions of \citet{madeira2018production}.
% By taking into account Aegaeon’s mass production rate, we can estimate the remaining lifetime of the G ring more realistically.

In addition, \citet{canup1997evolution} assumed that the G ring originates from the breakup of a progenitor satellite, and simulated the time evolution of the number of particles within the G ring. They found that the numerical simulation results with an assumption of a progenitor satellite with a radius of 1.5-$3\,\mathrm{km}$ match well with the observations.
Here we adopt this value to estimate the age of the G ring to be $10^{6}$-$10^{7}\,\mathrm{years}$ by dividing the total mass of the progenitor satellite by the $M^+$.

\section{Conclusions}
\label{Conclusions}
In this paper, the dynamical evolution of particles in the grain size of $[0.1, 50]\,{\mu}$m originating from the G-ring arc is studied. 
Various perturbation forces are considered in the dynamical model, including the effect of planetary oblateness, solar radiation pressure, Poynting-Roberson drag, Lorentz force, plasma drag and the gravity of moons.
In addition, the plasma sputtering is also taken into account.
We find that particles smaller than $1\,{\mu}$m leave the arc immediately, which are finally removed by the plasma sputtering.
The orbital distances of $1\,{\mu}$m particles oscillate with large amplitudes, which leads to the collision between particles and the A ring within a few years.
For particles with $r_\mathrm{g}>1\,{\mu}$m, they stay in the arc for a longer time compared to particles with $r_\mathrm{g}\leq1\,{\mu}$m.
These particles can leave the arc due to the plasma sputtering and plasma drag. Generally, dust particles escape from the arc with their grain sizes less than about $16\,{\mu}$m.
%Especially for large particles ($r_\mathrm{g}>5\,{\mu}$m), they are trapped in the arc until their grain sizes decay to nearly $5\,{\mu}$m regardless of their initial grain size.
After leaving the arc, particles move outward due to the plasma drag.
%When their grain sizes decay to below $\textcolor{red}{3}\,{\mu}$m, their orbital distances begin to oscillate and these particles hit a sink quickly.
Most of these particles finally collide with the A ring, while a small fraction of these particles hit Mimas or Aegaeon.
%Most of them collide with the A ring, while a \textcolor{red}{small} fraction of particles hit Mimas \textcolor{red}{and a smaller fraction of particles hit Aegaeon}.
Only particles with $r_\mathrm{g}\geq10\,{\mu}$m have opportunities to move outside the orbital distance of Enceladus.

Based on the study of the dynamics of dust particles, the formation and properties of the G ring are analyzed in this work. By comparing the normal $I/F$ of the simulated ring formed by particles originating from the G-ring arc and the number density near the G ring to the observations, we get the exponent $q=2.8$ for the particle size distribution of the G ring.
The good agreement in the normal $I/F$ between our result and the observation supports that the G ring may originate from the arc.
The observed normal $I/F$ in the part of the ring which passes through the arc is dominantly contributed by $[5, 20]\,{\mu}$m particles, while that in other parts of the ring are mainly contributed by $[5, 10]\,{\mu}$m particles.
The particles detected by the Cassini spacecraft at the radial distance of 2.7 $R_\mathrm{s}$ are mainly composed of dust particles smaller than $0.5\,{\mu}$m.
The peak value of the average particle number density of the G ring is on the order of $10^{-2}$.
The region of the arc is the brightest when viewed edge-on, where the value of the edge-on geometric optical depth is $3.9\times10^{-2}$.
If the observable threshold of the edge-on geometric optical depth is $1\times10^{-8}$, the maximum apparent edge-on thickness of the G ring is about $9,000\,\mathrm{km}$.
Based on our result of the particle size distribution of the G ring, the age of the G ring is estimated to be $10^{6}$ to $10^{7}\,\mathrm{years}$, and the remaining lifetime of the ring is on the order of $10^{4}\,\mathrm{years}$.

%Our simulation results 
%of the normal $I/F$ of the G ring and the number density near the G ring match well with the observations obtained by the Cassini spacecraft, which support other results, e.g., the particle properties, the apparent edge-on thickness, the age and the remaining lifetime in this work. 
 %match well with the observations obtained by the Cassini spacecraft, 
Our results of the properties of the G ring are robust since the relevant perturbation forces are considered and most of the parameters (such as the Saturn gravity field, the solar radiation pressure efficiency and the coefficient of Saturn’s magnetic field) used in our dynamical model are known with small uncertainties.
 %, and our simulation results match well with the observations obtained by the Cassini spacecraft.
The largest uncertainty of our results comes from that we only consider the water group ions $\mathrm{W}^+$ for the plasma drag in our dynamical model (see Section.~\ref{Dynamical model}), while the molecular oxygen ion $\mathrm{O}_2^+$ also contributes significantly to the plasma in the region of the G ring in the years when Saturn is close to the solstice \citep{elrod2012seasonal,elrod2014seasonal}.
% However, the plasma in the region of the G ring varies seasonally \citep{elrod2012seasonal,elrod2014seasonal}. 
%In the years when Saturn is close to the solstice, the molecular oxygen ion $\mathrm{O}_2^+$ also contributes significantly to the plasma in the region of the G ring.
%We have checked that the plasma drag becomes very large in these years if the $\mathrm{O}_2^+$ ion is also taken into account, and the dynamical evolution of dust particles changes significantly.
%As a result, the properties of the G ring which are determined by the dynamical evolution of dust particles are very different from the results in this paper.
This may indicate the G ring also varies seasonally due to the seasonal variation of the plasma in the vicinity of the G ring, which has not been mentioned in previous studies.
%and we will explode\textcolor{red}{explore} this consideration more deeply in future work.

\section*{Acknowledgements}

This work was supported by the National Natural Science Foundation of China (No.~12311530055, 12472048, and 12002397).

%%%%%%%%%%%%%%%%%%%%%%%%%%%%%%%%%%%%%%%%%%%%%%%%%%
\section*{Data Availability}
The basic data of this work will be shared on reasonable request to the corresponding author.
 
%The inclusion of a Data Availability Statement is a requirement for articles published in MNRAS. Data Availability Statements provide a standardised format for readers to understand the availability of data underlying the research results described in the article. The statement may refer to original data generated in the course of the study or to third-party data analysed in the article. The statement should describe and provide means of access, where possible, by linking to the data or providing the required accession numbers for the relevant databases or DOIs.



\bibliographystyle{mnras}
\bibliography{example} % if your bibtex file is called example.bib

%\begin{thebibliography}{99}
%\end{thebibliography}

%%%%%%%%%%%%%%%%%%%%%%%%%%%%%%%%%%%%%%%%%%%%%%%%%%

%%%%%%%%%%%%%%%%% APPENDICES %%%%%%%%%%%%%%%%%%%%%

%\appendix



%%%%%%%%%%%%%%%%%%%%%%%%%%%%%%%%%%%%%%%%%%%%%%%%%%

% Don't change these lines
\bsp	% typesetting comment
\label{lastpage}
\end{document}